\documentclass[a4paper,usenatbib,times,fleqn]{mn2e}
\usepackage{siunitx}
\usepackage{natbib,graphicx,amsmath,amssymb}
\usepackage{aas_macros,url,bm}
\usepackage{hyperref} 
	\hypersetup{colorlinks,citecolor=blue,linkcolor=blue,urlcolor=blue}
\usepackage[capitalise]{cleveref}
	\crefname{equation}{Equation}{Equations}
	\crefname{figure}{Figure}{Figures}
	\crefname{table}{Table}{Tables}
	
\usepackage{booktabs,times}
\usepackage{enumitem}

\usepackage{bm}
\newcommand*{\matrixbf}[1]{\bm{#1}}

\def\epsj{\epsilon_{j}}
\def\epsk{\epsilon_{k}}

\def\rhog{\rho_{\mathrm{g}}}
\def\rhod{\rho_{\mathrm{d}}}
\def\rhodj{\rho_{\mathrm{d}j}}
\def\rhogO{\rho_{\mathrm{g},0}}

\def\rhograin{\rho_{\rm grain}}
\def\rhoeff{\rho_\mathrm{eff}}

\def\sj{s_j}
\def\smin{s_{\rm min}}
\def\smax{s_{\rm max}}

\def\sumj{\sum_{j}}
\def\sumk{\sum_{k}}

\def\vb{\textbf{v}}
\def\vg{\textbf{v}_{\mathrm{g}}}
\def\vd{\textbf{v}_{\mathrm{d}}}
\def\vdone{v_{\mathrm{d}}}
\def\vdj{\textbf{v}_{\mathrm{d}j}}

\def\vK{v_\mathrm{K}}
\def\OmegaK{\Omega_\mathrm{K}}

\def\cs{c_\mathrm{s}}

\def\deltav{\Delta \textbf{v}}
\def\deltavj{\Delta \textbf{v}_{j}}

\def\f{\mathbf{f}}
\def\fg{\mathbf{f}_{\mathrm{g}}}
\def\fdj{\mathbf{f}_{\mathrm{d}j}}

\def\deltafj{\Delta \mathbf{f}_{j}}
\def\deltafk{\Delta \mathbf{f}_{k}}

\def\tj{t_{j}}
\def\tk{t_{k}}
\def\tsone{t_{\mathrm{s}}^{N=1}}
\def\Ts{T_{\mathrm{s}}}
\def\Tsj{T_{\mathrm{s}j}}
\def\Tsk{T_{\mathrm{s}k}}
\def\tTsj{\widetilde{T}_{\mathrm{s}j}}

\def\tcfl{t_\text{\tiny CFL}}

\def\barrhod{\bar{\rho}_\mathrm{d}}

\def\tildevd{\tilde{v}_\mathrm{d}}
\def\barvd{\bar{v}_\mathrm{d}}

\def\barz{\bar{z}}
\def\bart{\bar{t}}
\def\St{\mathrm{St}}

\def\dst{\displaystyle}

\def\pasa{PASA}

\defcitealias{Laibe/Price/2011}{LP11}
\defcitealias{Laibe/Price/2012}{LP12a}
\defcitealias{Laibe/Price/2012a}{LP12b}
\defcitealias{Laibe/Price/2014a}{LP14a}
\defcitealias{Laibe/Price/2014b}{LP14b}
\defcitealias{Laibe/Price/2014c}{LP14c}
\defcitealias{Price/Laibe/2015}{PL15}

\title[Multiple grain sizes in SPH]{MULTIGRAIN: A smoothed particle hydrodynamics algorithm for multiple small dust grains and gas}

\author[Hutchison, Price \& Laibe]{Mark Hutchison$^{1,2}$\thanks{markahutch@gmail.com}, Daniel J. Price$^{3}$, and Guillaume Laibe$^{4}$  \\
$^{1}$Physikalisches Institut, Universit{\"a}t Bern, Gesellschaftstrasse 6, 3012 Bern, Switzerland \\
$^{2}$Institute for Computational Science, University of Zurich, Winterthurerstrasse 190, CH-8057 Z{\"u}rich, Switzerland \\
$^{3}$Monash Centre for Astrophysics and School of Physics \& Astronomy, Monash University, Clayton, Vic 3800, Australia \\
$^{4}$Univ Lyon, Univ Lyon1, Ens de Lyon, CNRS, Centre de Recherche Astrophysique de Lyon UMR5574, F-69230, Saint-Genis-Laval, France
}
\pagerange{\pageref{firstpage}--\pageref{lastpage}} \pubyear{2017}

\begin{document}
\label{firstpage}
\bibliographystyle{mnras}
\maketitle

\begin{abstract}
We present a new algorithm, \textsc{multigrain}, for modelling the dynamics of an entire population of small dust grains immersed in gas, typical of conditions that are found in molecular clouds and protoplanetary discs. The \textsc{multigrain} method is more accurate than single-phase simulations because the gas experiences a backreaction from each dust phase and communicates this change to the other phases, thereby indirectly coupling the dust phases together. The \textsc{multigrain} method is fast, explicit and low storage, requiring only an array of dust fractions and their derivatives defined for each resolution element.
\end{abstract}



\section{Introduction}
\label{sec:physics}
Modelling the interaction of multiple dust grains simultaneously with the gas is a `grand challenge' in protoplanetary disc modelling \citep{Haworth/etal/2016}, since discs involve grains with sizes spanning several orders of magnitude, from sub-micron grains to km-sized planetesimals. Grains of different sizes experience different dynamics since small grains are lighter and more easily influenced by the gas compared to larger, heavier grains.

The usual approach to dusty gas dynamics is to model the gas and dust as separate fluids. The gas is modelled either on a grid \citep{Paardekooper/Mellema/2004,Youdin/Johansen/2007,Balsara/etal/2009,Bai/Stone/2010a,Miniati/2010,Yang/Johansen/2016} or on a set of Lagrangian particles \citep{Monaghan/Kocharyan/1995,Monaghan/1997b,Barriere-Fouchet/etal/2005,Laibe/Price/2012a,Laibe/Price/2012b,Loren-Aguilar/Bate/2014,Loren-Aguilar/Bate/2015a}; similarly for each dust phase (although the discretisation method often defaults to the one used by the gas). During simulation, the gas and dust fluids are evolved independently, but interact via a common drag force \citep[e.g.][]{Saffman/1962,Garaud/Lin/2004}.

Although grid- and particle-based methods each have their own distinct advantages/disadvantages \citep[e.g.][]{Price/Federrath/2010}, they both require prohibitively small timesteps or implicit methods at high drag. Furthermore, \citet{Laibe/Price/2012a,Laibe/Price/2012b} discovered a drag resolution criterion that becomes increasingly restrictive with smaller grain sizes and applies generally to any method that models dust on a grid or on a set of particles that is not colocated with the gas at all times. While \citet{Laibe/Price/2012a,Laibe/Price/2012b}, and later \citet{Loren-Aguilar/Bate/2014}, tested this spatial criterion using smoothed particle hydrodynamics (SPH), \citet{Youdin/Johansen/2007} inferred a similarly high resolution requirement in hybrid grid-particle simulations. Failing to meet this criterion may explain the first-order convergence rate in high drag regimes observed by \citet{Miniati/2010}, \citet{Bai/Stone/2010a}, and \citet{Yang/Johansen/2016}.

To address the restrictive temporal and spatial restrictions that exist for high drag regimes,  \citet{Laibe/Price/2014a,Laibe/Price/2014b,Laibe/Price/2014c} (hereafter \citetalias{Laibe/Price/2014a,Laibe/Price/2014b,Laibe/Price/2014c}) and \citet{Price/Laibe/2015} (hereafter \citetalias{Price/Laibe/2015}) developed a single-fluid formulation appropriate for small grains --- similar to earlier formulations by \citet{Johansen/Klahr/2005}. The dust-gas mixture is advected at the barycentric velocity and whose density is equal to the total density of the mixture. In the context of smoothed particle hydrodynamics, this means the mixture is represented by a single set of SPH particles with an evolution equation for the dust fraction \citepalias{Laibe/Price/2014b,Price/Laibe/2015}.

While the above methods provide a means of modelling discs or molecular clouds with a single embedded dust phase, the challenge is to span the observed range of grain sizes. The typical approach is the one we recently used in \citet{Dipierro/etal/2015}, where a series of single-phase simulations were stitched together in post-processing to interpret the dark structures observed at millimetre wavelengths by the ALMA interferometer in the disc surrounding the star HL Tau. In that paper, the method from \citet{Laibe/Price/2012a} was used to model the dynamics of mm-sized grains and larger, while the smaller grains were modelled using the method from \citetalias{Price/Laibe/2015}. Besides being tedious, the procedure used by \citet{Dipierro/etal/2015} is slow and, more importantly, neglects the indirect coupling between dust phases caused by the `backreaction' of individual phases on the gas, which in turn influences the grain dynamics. Neglecting this backreaction misses important effects such as outwards migration of dust particles \citep{Bai/Stone/2010a} and/or modification of the linear growth rate of the streaming instability \citepalias{Laibe/Price/2014c}. Also, backreaction in individual grain size simulations is both annoying and wrong --- annoying because the different response of the gas makes stacking of different dust grain distributions difficult \citep{Tricco/Price/Laibe/2017}; wrong because the gas should respond to the entire dust mixture, rather than each grain size individually.
  
 In the present paper, we develop a new \textsc{multigrain} algorithm for modelling the dynamics of multiple dust phases, based on the analytical work presented in \citetalias{Laibe/Price/2014c}. Because much of the opacity and accompanying scattering/emission in astrophysical environments stems from the presence of dust grains that can be considered `small' (i.e. where the terminal velocity approximation is valid), we focus on deriving and implementing the SPH versions of the continuum equations for the multi-phase, terminal velocity approximation --- generalising the single-dust-phase method developed in \citetalias{Price/Laibe/2015}.

\section{The diffusion approximation for multiple dust species}
\label{sec:diffusion_approximation}

\subsection{Continuum equations}
\label{sec:continuum_equations}

We consider a system consisting of a mixture of a single gas phase and $N$ strongly-coupled dust phases. Throughout this paper, we use the indices $a$, $b$, and $c$ to refer to individual simulation particles that move at the barycentric velocity of the mixture. Subscript or superscripts $\text{g}$ and $\text{d}$ are used for gas and dust properties, respectively. Finally, we identify the fluid quantities for each of the $N$ different dust phases using the index $j$.

\subsubsection{General equations}
\label{sec:general_equations}

\citetalias{Laibe/Price/2014c} derived the general continuum fluid equations for a mixture of gas and $N$ coupled dust species moving in a barycentric reference frame. They further showed that in strongly-coupled regimes --- i.e. first order in $\tj/T$, where $\tj$ is a drag timescale specific to each grain type (see \namecref{eq:fluideqs_drag_time}~\ref{eq:fluideqs_drag_time} below; note the difference in notation from that of \citetalias{Laibe/Price/2014c}) and $T$ is the timescale for a sound wave to propagate over a typical distance L \citep[commonly referred to as the terminal velocity approximation; see e.g.][]{Youdin/Goodman/2005,Chiang/2008,Barranco/2009,Lee/etal/2010,Jacquet/Balbus/Latter/2011} --- the fluid equations reduce to
\begin{align}
	\frac{{\rm d} \rho}{{\rm d} t}& =  - \rho \left( \nabla \cdot \vb \right), 
	\label{eq:drhodt}
\\
	\frac{{\rm d} \epsj}{{\rm d} t} & =  - \frac{1}{\rho} \nabla \cdot 
		\left[ \rho \epsj \left( \deltavj - \epsilon \deltav \right) \right],
	\label{eq:depsdt} 
\\
	\frac{{\rm d} \vb}{{\rm d} t} & = \left( 1 - \epsilon \right) \fg + \sumj \epsj \fdj + \f,
	\label{eq:dvdt} 
\\
	\frac{{\rm d} u}{{\rm d} t} & = - \frac{P}{\rhog} \nabla \cdot \vb + \epsilon \deltav \cdot \nabla u,
	\label{eq:dudt}
\\
	\deltavj & = \left[ \deltafj - \sumk \epsk \deltafk \right] \epsj \tj,
	\label{eq:def_deltavj}
\end{align}
where $\mathrm{d}/\mathrm{d}t$ is the convective derivative using the barycentric velocity $\vb$,
\begin{equation}
	\vb \equiv \dst \frac{\rhog \vg + \dst \sumj \rhodj \vdj }{\rho} = \frac{\rhog \vg + \rhod \vd}{\rho},
	\label{eq:def_v}
\end{equation}
$\rho$ is the total density of the mixture,
\begin{equation}
	\rho \equiv \rhog + \rhod  = \rhog + \sumj \rhodj,
	\label{eq:def_rho}
\end{equation}
$\epsj$ and $\epsilon$ are the mass fractions (relative to the mixture) of the individual and combined dust phases, respectively,
\begin{align}
	\epsj & \equiv \frac{\rhodj}{\rho},
	\label{eq:def_epsj}
\\
	\epsilon & \equiv \sumj \epsj = \frac{\rhod}{\rho},
	\label{eq:def_epsilon}
\end{align}
$\deltav$ is the weighted sum of the differential velocities $\deltavj \equiv \vdj-\vg$,
\begin{equation}
	\deltav \equiv \frac{1}{\epsilon} \sumj \epsj \deltavj,
	\label{eq:def_deltav}
\end{equation}
$\f$ represents accelerations acting on both components of the fluid while $\fg$ and $\fdj$ represent the accelerations acting on the gas and dust components, respectively, $\deltafj \equiv \fdj - \fg$ is the differential force between the gas and each dust phase, $u$ is the specific thermal energy of the gas, and $P$ is the gas pressure.

\subsubsection{Drag timescales}
\label{sec:drag_timescales}

When $N = 1$, the drag timescale is unambiguously set by the drag stopping time,
\begin{equation}
	\tsone \equiv \frac{\rhog \rhod}{K \rho},
	\label{eq:def_Ts_N=1}
\end{equation}
where $K$ is a drag coefficient that, in general, depends on local properties of the gas and dust. We assume that $K$ is either constant or in the linear Epstein regime, suitable for small dust grains with low Mach numbers \citep[][also e.g. \citealt{Laibe/Price/2012b}]{Epstein/1924}. In the latter case, 
\begin{equation}
	K = \frac{\rhog \rhod}{\rhograin s} \sqrt{\frac{8}{\pi \gamma}} \cs = \frac{\rhog \rhod \cs}{\rhoeff s},
	\label{eq:drag_constant}
\end{equation}
where we assume spherical grains with radius $s$, with uniform intrinsic dust density $\rhograin$, or equivalently, an effective density $\rhoeff \equiv \rhograin \sqrt{\pi \gamma / 8}$. As usual, $\gamma$ is the adiabatic constant. The stopping time for $N=1$ in the Epstein regime can therefore be written as
\begin{equation}
	\tsone =  \frac{\rhoeff s}{\rho \cs}.
	\label{eq:stopping_time}
\end{equation}

Generalising the stopping time to $N>1$ is conceptually simple, but difficult in practice. Each dust type equilibrates with the gas at a different rate depending on both the intrinsic properties of the dust grains and the local properties of the gas. Although we assume dust grains of different species do not interact, they are indirectly coupled by their mutual backreaction on the gas. One approach is to derive timescales using the eigenvalues of the drag matrix \citepalias{Laibe/Price/2014c}, but the derivations and the expressions become increasingly unwieldy as $N$ increases (i.e. there is no general algebraic expression as a function of $N$).

The eigenvalues help aid in interpreting results, but they are not needed to evolve the fluid equations numerically. The only potential impact the eigenvalues have is through their influence on the timestep. Even then, \citetalias{Laibe/Price/2014c} found fixed upper/lower bounds to the eigenvalues of the $N \times N$ drag matrix, effectively removing any need for the eigenvalues during computation.

For convenience, we define the following timescales to help simplify our numerical implementation:
\begin{alignat}{2}
	\Tsj & \equiv \frac{\rhoeff \sj}{\rho \cs}  && = \epsj \left( 1 - \epsilon \right) \tj, 
	\label{eq:def_Tsj}
\\
	\tTsj & \equiv \frac{\Tsj - \sumk \epsk \Tsk}{1 - \epsilon} \quad && = \epsj \tj - \sumk \epsk^2 \tk, 
	\label{eq:def_tTsj}
\end{alignat}
where
\begin{equation}
	\tj \equiv \frac{\rho}{K_j},
	\label{eq:fluideqs_drag_time}
\end{equation}
and where $K_j$ is the drag coefficient for each dust phase, e.g.
\begin{equation}
	K_j =  \frac{\rhog \rhodj \cs}{\rhoeff \sj}.
	\label{eq:drag_constant_j}
\end{equation}
Note that the weighted sums of \cref{eq:def_Tsj,eq:def_tTsj} happen to be equivalent, i.e. 
\begin{equation}
	\frac{1}{\epsilon} \sumj \epsj \Tsj  \; = \; \frac{1}{\epsilon} \sumj \epsj \tTsj 
		\; = \; \frac{1 - \epsilon}{\epsilon} \sumj \epsj^2 \tj.
	\label{eq:equiv_sums}
\end{equation}
This new quantity carries physical significance, but its interpretation is clearer if we first define an effective grain size for the mixture,
\begin{equation}
	s \equiv \frac{1}{\epsilon} \sumj \epsj \sj,
	\label{eq:def_seff}
\end{equation}
such that \cref{eq:equiv_sums} can be written in a more familiar form:
\begin{equation}
	\Ts \equiv \frac{1}{\epsilon} \sumj \epsj \Tsj = \frac{\rhoeff s}{\rho \cs}.
	\label{eq:def_Ts}
\end{equation}
Comparing this to \cref{eq:stopping_time}, one may observe that $\Ts$ acts like an effective stopping time for the mixture.

The benefit of using $\Tsj$ and $\tTsj$ in lieu of $\tj$ is that they allow us to use our existing codebase with only a few additional lines of code, namely to assemble $\tTsj$ ($\Tsj$ is calculated identically to $\tsone$ with $s$ replaced by $\sj$). In return, the form of the evolution equations are unchanged from the $N = 1$ case, as evidenced in the following sections.

\subsubsection{Hydrodynamics}
\label{sec:hydrodynamics}

For the simple case of hydrodynamics, the only force is the pressure gradient, i.e.
\begin{align}
	\fdj & = 0,
\\
	\fg & = - \frac{\nabla P}{\rhog},
	\label{eq:fg_simp}
\\
	\deltafj & = \frac{\nabla P}{\rhog}.
	\label{eq:deltaf_simp}
\end{align}
Using \cref{eq:def_Tsj,eq:deltaf_simp} to simplify \cref{eq:def_deltavj}, we get
\begin{equation}
	\deltavj = \frac{\epsj \tj \nabla P}{\rho} = \frac{\Tsj \nabla P}{\rhog},
	\label{eq:deltavj_simp}
\end{equation}
while \cref{eq:def_deltav,eq:def_Ts,eq:deltavj_simp} allow us to write
\begin{equation}
	\deltav = \frac{\Ts \nabla P}{\rhog}.
	\label{eq:deltav_simp}
\end{equation}
As promised, when these last two expressions for $\deltavj$ and $\deltav$ are inserted into \cref{eq:drhodt,eq:depsdt,eq:dvdt,eq:dudt}, we obtain the same form of the fluid equations as reported in \citetalias{Price/Laibe/2015} for the $N = 1$ terminal velocity approximation, namely
\begin{align}
	\frac{{\rm d} \rho}{{\rm d} t}& =  - \rho \left( \nabla \cdot \vb \right), 
	\label{eq:drhodt_simp}
\\
	\frac{{\rm d} \epsj}{{\rm d} t} & =  - \frac{1}{\rho} \nabla \cdot 
		\left( \epsj \tTsj \nabla P \right),
	\label{eq:depsdt_simp} 
\\
	\frac{{\rm d} \vb}{{\rm d} t} & = -\frac{\nabla P}{\rho} + \f,
	\label{eq:dvdt_simp} 
\\
	\frac{{\rm d} \tilde{u}}{{\rm d} t} & = - \frac{P}{\rho} \nabla \cdot \vb.
	\label{eq:dudt_simp}
\end{align}
where for convenience we have defined $\tilde{u} \equiv (1 - \epsilon) u$ instead of evolving $u$ directly as in \citetalias{Price/Laibe/2015}. The corresponding energy equation in terms of $u$ would be
\begin{equation}
	\frac{{\rm d} u}{{\rm d} t} = - \frac{P}{\rhog} \nabla \cdot \vb + \frac{\epsilon \Ts}{\rhog} \nabla P \cdot \nabla u.
\end{equation}

In order to recover the special case of a single dust phase, we need only collapse the sums in \cref{eq:def_Tsj,eq:def_tTsj} and set $\sj \to s$ and $\epsj \to \epsilon$. It is simple to check that in this limit, $\Tsj = \tTsj = \Ts = \tsone$, thereby recovering the $N = 1$ fluid equations from \citetalias{Price/Laibe/2015} exactly. 

\subsubsection{Equation of state}
\label{sec:equation_of_state}

The set of equations above is closed by assuming the usual equation of state, which constrains the gas pressure $P$ in terms of the gas density and temperature. Unless otherwise specified in this paper, we assume an adiabatic equation of state, i.e.
\begin{equation}
	P = \left( \gamma - 1 \right) \rhog u  = \left( \gamma -1 \right) \left( 1 - \epsilon \right) \rho u,
	\label{eq:eq_of_state}
\end{equation}
or simply
\begin{equation}
P = (\gamma -1) \rho \tilde{u}.
\end{equation}

\subsection{Timestepping}
\label{sec:timestepping}

As pointed out by \citetalias{Price/Laibe/2015}, the addition of the dust evolution equation adds a further constraint on the timestep that becomes limiting when the diffusion coefficient is large. We can derive this timestep constraint more rigorously than that presented by \citetalias{Price/Laibe/2015}, albeit with the same result for $N=1$, by discretising the set of equations in time using a forward Euler method
\begin{align}
	\frac{\rho^{n+1} - \rho^n}{\Delta t}& =  - \rho \left( \nabla \cdot \vb \right), 
	\label{eq:drhodt_simpdt}
\\
	\frac{\epsilon_j^{n+1} - \epsilon_j^{n}}{\Delta t} & =  - \frac{1}{\rho} \nabla \cdot 
		\left( \epsj \tTsj \nabla P \right),
	\label{eq:depsdt_simpdt} 
\\
	\frac{\bm{v}^{n+1} - \bm{v}^n}{\Delta t} & = -\frac{\nabla P}{\rho},
	\label{eq:dvdt_simpdt} 
\end{align}
and performing a Von Neumann stability analysis on the above semi-discrete equations. That is, we solve the linear system that results from assuming plane wave solutions of the form
\begin{align}
	\rho & = D e^{i (\bm{k}\cdot\bm{x} - \omega t)},
\\
	\bm{v} & = \bm{V} e^{i (\bm{k}\cdot\bm{x} - \omega t)},
\\
	\epsilon_j & = E_j e^{i (\bm{k}\cdot\bm{x} - \omega t)},
\end{align}
where $D$, $\bm{V}$, and $E_j$ are perturbation amplitudes, $k$ is the wave number, $\bm{x}$ is the position vector, and $\omega$ is the angular frequency. This analysis generically produces a timestep criterion of the form
\begin{equation}
	\Delta t < C_0\frac{1}{k c_{\max}},
\end{equation}
where $C_0$ is a dimensionless safety factor of order unity and $c_{\max}$ is the maximum wave speed according to the dispersion relation for linear waves. The wavelength of maximum growth usually occurs on the resolution scale, giving the usual Courant criterion
\begin{equation}
	\Delta t < C_0\frac{h}{c_{\max}},
\end{equation}
where $h$ is the SPH smoothing length. For $N=1$ the dispersion relation to first order in $\omega t_{\rm s}$ is given by \citep{Laibe/Price/2014a}
\begin{equation}
	\omega = \pm \tilde{c}_{\rm s} k - \frac{i}{2} t_{\rm s} k^2 c_{\rm s}^2 \epsilon,
\end{equation}
where $\tilde{c}^2_{\rm s} \equiv c_{\rm s}^2 (1 - \epsilon)$ is the modified sound speed (squared). The maximum wave speed is therefore
\begin{equation}
	c_{\max} =\left\vert \frac{\omega}{k}\right\vert = \sqrt{\tilde{c}_{\rm s}^2 + \frac{1}{4} \epsilon^2 t_{\rm s}^2 k^2 c_{\rm s}^4}.
\end{equation}
and the timestep constraint appropriate for SPH is
\begin{equation}
	\Delta t < C_0\frac{h}{\sqrt{\tilde{c}^2_{\rm s} + \epsilon^2 t^2_{\rm s} c_{\rm s}^4 / h^2}}.
\end{equation}
This is similar to the timestep criterion proposed by  \citetalias{Price/Laibe/2015} except that the above combines the usual Courant-Friedrichs-Lewy (CFL) condition ($\Delta t < h/\tilde{c}_{\rm s}$) and the additional constraint from the dust evolution ($\Delta t < h^2 / (\epsilon t_{\rm s} c_{\rm s}^2)$) into a single criterion.

When generalising to multiple dust phases, we find the same result but with the effective stopping time replacing the $N=1$ stopping time, giving
\begin{equation}
	\Delta t < C_0\frac{h}{\sqrt{\tilde{c}^2_{\rm s} + \epsilon^2 T_{\rm s}^2 c_{\rm s}^4 / h^2}}.
	\label{eq:deltat_eps}
\end{equation}

As expected, with $\Ts$ in the denominator, restricting ourselves to strong drag regimes \textit{weakens} the constraint on the timestep. More specifically, the timestep is limited when the grain-size distribution is dominated by large grains (or, alternatively, high dust fraction), such that
\begin{equation}
	\frac{\epsilon \Ts}{1-\epsilon} > \Delta \tcfl,
\end{equation}
where $\Delta \tcfl \equiv h/\tilde{c}_{\rm s}$ is the CFL timestep. The added advantage of the criterion in \cref{eq:deltat_eps} is that it is less stringent than the explicit timestep for either the full \textsc{multigrain} one-fluid formalism or the multi-fluid method (Equation~79 and 80 of \citetalias{Laibe/Price/2014c}, respectively):
\begin{align}
	\Delta t_{\mathrm{one-fluid}} & < C \left[  \max_{j}\left(\frac{1}{\epsj \tj} \right) +
		\frac{1}{\left(1 - \epsilon \right)} \sumj \tj^{-1}  \right]^{-1},
	\label{eq:deltat_one}
\\
	\Delta t_{\mathrm{multi-fluid}} & < C \min_{j} \left[  \frac{1}{\tj} \left( \frac{1}{\epsj} + 
		\frac{1}{1-\epsilon} \right) \right]^{-1},
	\label{eq:deltat_multi}
\end{align}
where $C$ is another safety factor. Thus, as long as the cut-off to our dust distribution is $\lesssim \mathrm{cm}$ \citepalias[see][]{Price/Laibe/2015}, our global timestep should be of the order of $\Delta \tcfl$.

\section{SPH formulation}
\label{sec:SPH_formulation}

When formulating the discretised SPH fluid equations, we can take advantage of the fact that (i) the only equations that were altered by having multiple dust phases were the dust fraction and energy equations and (ii) we have written the continuum equations in the same form as \citetalias{Price/Laibe/2015}.

The first point allows us to adopt the discretised density and momentum equations from \citetalias{Price/Laibe/2015} without any changes (thereby guaranteeing exact conservation of linear and angular momentum),
\begin{align}
	\rho_{a} & =  \sum_{b} m_{b} W_{ab} (h_{a}),
	\label{eq:sph_density}
\\
	\frac{{\rm d}\bm{v}_a}{{\rm d}t} & = -\sum_{b} m_{b} \left[ \frac{P_{a} + 
		q^{\rm AV}_{ab, a}}{\Omega_{a} \rho_{a}^{2}} \nabla_{a} W_{ab} (h_{a}) + \right.  \nonumber
	\\
		& \phantom{{}= -\sum_{b} m_{b} \left[ \right.} \left. \frac{P_{b} + q^{\rm AV}_{ab, b}}{\Omega_{b} \rho_{b}^{2}} \nabla_{a} W_{ab} (h_{b}) \right] + \bm{f}_a,
	\label{eq:sph_momentum}
\end{align}
where $W_{ab}$ is the usual SPH kernel, $h$ is the smoothing length, $\Omega$ is the usual term to account for smoothing length gradients
\begin{equation}
	\Omega_{a} = 1 - \frac{\partial h_{a}}{\partial \rho_{a}} \sum_{b} m_{b} \frac{\partial W_{ab} (h_{a})}{\partial h_{a}},
\end{equation}
and $h$ is related to $\rho$ in the usual manner (which requires an iterative procedure to solve \cref{eq:sph_density}; see \citealt{Price/Monaghan/2004,Price/Monaghan/2007}, \citetalias{Laibe/Price/2014b}).

The second point allows us to write down the generalised diffusion equation for $\epsj$ by inspection. Comparing \cref{eq:depsdt_simp} to equation~12 in \citetalias{Price/Laibe/2015} suggests that we can use either of their discretised diffusion equations provided we make the substitutions $\tsone \to \tTsj$ and $\epsilon \to \epsj$ (although in the latter case, care must be taken to leave any instances of the gas fraction, $1-\epsilon$, untouched). Furthermore, because evolving the dust fraction directly can in some instances result in negative values, we prefer to use the positive definite formulation prescribed in Appendix~B of \citetalias{Price/Laibe/2015} by defining $S_j \equiv \sqrt{\rho \epsj} $ (not to be confused with the grain size $s_j$). The corresponding evolution equation in terms of $S_j$ is
\begin{align}
	\frac{\mathrm{d} S_{j,a}}{\mathrm{d}t} &  =   - \frac{1}{2} \sum_{b} \frac{m_{b} S_{j,b}}{\rho_{b}} 
		\left( \frac{\widetilde{T}_{\mathrm{s}j,a}}{\rho_{a}} + \frac{\widetilde{T}_{\mathrm{s}j,b}}{\rho_{b}} \right)
		\left(P_{a} - P_{b} \right) \frac{\overline{F}_{ab}}{\vert r_{ab} \vert} \nonumber
	\\
		& \phantom{{}=} + \frac{S_{j,a}}{2\rho_{a}\Omega_{a}} \sum_{b} m_{b} \bm{v}_{ab}
		\cdot \nabla_a W_{ab} (h_{a}),
	\label{eq:sph_dustfrac}
\end{align}
where $\overline{F}_{ab} \equiv \frac{1}{2}[ F_{ab}(h_{a}) + F_{ab}(h_{b}) ]$ and $F_{ab}$ is defined such that $\nabla_{a} W_{ab} \equiv F_{ab} \hat{\bm{r}}_{ab}$. In writing the diffusion equation in this form, we have implicitly chosen to use the faster, easier-to-implement `direct second derivative' method; however, the evolution equation for the `two first derivatives' method can be obtained in the same fashion (see \citetalias{Price/Laibe/2015} for a comparison of these two methods).

\subsection{Conservation of energy}
\label{sec:conservation_of_energy}

This leaves only the energy equation to be determined. It is tempting to simply generalise the equation for the energy in a similar manner to the above, but conservation of energy puts an additional constraint on the form of the equation that is not immediately obvious. Instead, we derive the energy equation using the already discretised fluid equations above and by enforcing exact conservation of energy.

The total energy $E$ of the system in the terminal velocity approximation can be expressed as
\begin{equation}
	E = \sum_a m_a \left( \frac{1}{2} v_a^2 + \tilde{u}_a \right),
	\label{eq:total_energy}
\end{equation}
where $\tilde{u}_a \equiv (1 - \epsilon_a) u_a$ as previously. Conservation of energy requires that
\begin{align}
	\frac{\mathrm{d} E}{\mathrm{d}t} & = \sum_{a} m_{a} \left[ \bm{v}_{a} 
		\cdot \frac{\mathrm{d} \bm{v}_{a}}{\mathrm{d} t} + 
	        \frac{\mathrm{d} \tilde{u}_{a}}{\mathrm{d} t} \right]= 0.
\end{align}
where
\begin{equation}
 \frac{\mathrm{d} \tilde{u}_{a}}{\mathrm{d} t} = 
		\frac{\rho^\mathrm{g}_{a}}{\rho_a} \frac{\mathrm{d} u_{a}}{\mathrm{d} t} - u_{a} \sumj 
		\left( \frac{2 S_{j,a}}{\rho_{a}} \frac{\mathrm{d} S_{j,a}}{\mathrm{d} t} - 
		\frac{S^2_{j,a}}{\rho^2_{a}} \frac{\mathrm{d} \rho_{a}}{\mathrm{d}t} \right).
\end{equation}

Inserting the different expressions from \cref{eq:sph_density,eq:sph_momentum,eq:sph_dustfrac} and solving for the time derivative of the energy dictates that the discretised energy equation should be
\begin{equation}
	\frac{{\rm d}\tilde{u}_{a}}{{\rm d}t} = \sum_{b} m_{b} \frac{P_{a} + 
		q^{\rm AV}_{ab, a}}{\Omega_{a} \rho_{a}^{2}}  \left( \bm{v}_{a} - \bm{v}_{b}\right) \cdot
		\nabla_{a} W_{ab} (h_{a}),
	\label{eq:sph_energy1}
\end{equation}
or, if one evolves $u$ directly as in \citetalias{Price/Laibe/2015}
\begin{align}
	\frac{{\rm d}u_{a}}{{\rm d}t} & =  \frac{1}{1 - \epsilon_{a}} \sum_{b} m_{b} \frac{P_{a} + 
		q^{\rm AV}_{ab, a}}{\Omega_{a} \rho_{a}^{2}}  \left( \bm{v}_{a} - \bm{v}_{b}\right) 
		\cdot \nabla_{a} W_{ab} (h_{a}) \nonumber 
	\\
		& \phantom{{} =} - \frac{\rho_{a}}{2 \rho^\mathrm{g}_{a}} \sumj \sum_{b} m_{b} 
		\frac{S_{j,a} S_{j,b}}{\rho_{a} \rho_{b}} \left( \frac{\widetilde{T}_{\mathrm{s}j,a}}{\rho_{a}} + 
		\frac{\widetilde{T}_{\mathrm{s}j,b}}{\rho_{b}} \right) \nonumber
	\\
		& \phantom{{} = - \frac{\rho_{a}}{2 \rho^\mathrm{g}_{a}} \sumj \sum_{b}} \left( u_{a} - u_{b}\right) 
		\left( P_{a} - P_{b} \right) \frac{\overline{F}_{ab}}{\vert r_{ab} \vert}.
	\label{eq:sph_energy}
\end{align}

\subsection{Shock-capturing terms}
\label{sec:shock_terms}

We include the artificial viscosity and conductive terms below for completeness, but note that they are unchanged by the addition of more dust phases.

\subsubsection{Artificial viscosity}

The artificial viscosity term is computed as follows:
\begin{equation}
	q^{\rm AV}_{ab, a} =
	\begin{cases}
		-\frac{1}{2} \left( 1 - \epsilon_{a} \right) v_{\mathrm{sig},a} \bm{v}_{ab} 
			\cdot \hat{\bm{r}}_{ab}, & \qquad \bm{v}_{ab} \cdot \hat{\bm{r}}_{ab} < 0
	\\	
		0,  & \qquad \mathrm{otherwise},
	\end{cases}
\end{equation}
where $\bm{v}_{ab} \equiv \bm{v}_{a} - \bm{v}_{b}$ (similarly for $\hat{\bm{r}}_{ab}$) and the signal speed $v_{\mathrm{sig}}$ corresponds to the usual choice for hydrodynamics, i.e.
\begin{equation}
	v_{\mathrm{sig},a} = \alpha^{\rm AV}_{a} c_{\mathrm{s},a} + \beta^{\rm AV} \vert \bm{v}_{ab} \cdot \hat{\bm{r}}_{ab} \vert,
\end{equation}
where $\alpha^{\rm AV}_{a} \in [0,1]$ is the linear dimensionless viscosity parameter \citep[the index implying that $\alpha^{\rm AV}$ can be unique to each particle; see, e.g.,][]{Morris/Monaghan/1997,Cullen/Dehnen/2010} and $\beta^{\rm AV}$ (typically $\beta^{\rm AV} = 2$) is the von Neumann-Richtmyer viscosity parameter.

\subsubsection{Artificial conductivity}

In order to correctly treat contact discontinuities, an artificial conductivity term must be added to the energy equations \citep[see][]{Price/2008},
\begin{equation}
	\left( \frac{\mathrm{d} u_{a}}{\mathrm{d} t} \right)_{\rm cond} = \frac{1}{1-\epsilon_{a}} \sum_{b} m_{b}
		\left[ \frac{Q_{ab, a}}{\Omega_{a} \rho_{a}^{2}} F_{ab} (h_{a}) + 
		\frac{Q_{ab, b}}{\Omega_{b} \rho_{b}^{2}} F_{ab} (h_{b}) \right],
\end{equation}
where
\begin{equation}
	Q_{ab,a} = \frac{1}{2} \alpha_u \rho_{a} v_{\mathrm{sig},u} \left( u_{a} - u_{b} \right),
\end{equation}
with $\alpha_u \in [0,1]$ the dimensionless conductivity parameter and $v_{\mathrm{sig},u} = \vert \bm{v}_{ab} \cdot \hat{\bm{r}}_{ab} \vert$ \citep{Price/2008,Wadsley/Veeravalli/Couchman/2008}.

\section{Numerical tests}
\label{sec:numerical_tests}

Given the similarity of the SPH equations in \cref{sec:SPH_formulation} to those in \citetalias{Price/Laibe/2015} and the existing implementation of the latter in our SPH code \textsc{phantom} \citep[e.g.][]{Dipierro/etal/2015,Price/etal/2017}, the generalisation to $N$ dust phases was straightforward. \textsc{Phantom} is well tested \citep[see][]{Price/etal/2017} and we are confident that the implementation of the $N=1$ terminal velocity approximation from which we started was correct. Therefore, the tests in this section are less focused on the code as a whole and more focused on specific aspects of our implementation.

\subsection{Recovering the N=1 case}
\label{sec:recoveringN1}

\begin{figure*}
	\centering{\includegraphics[width=\textwidth]{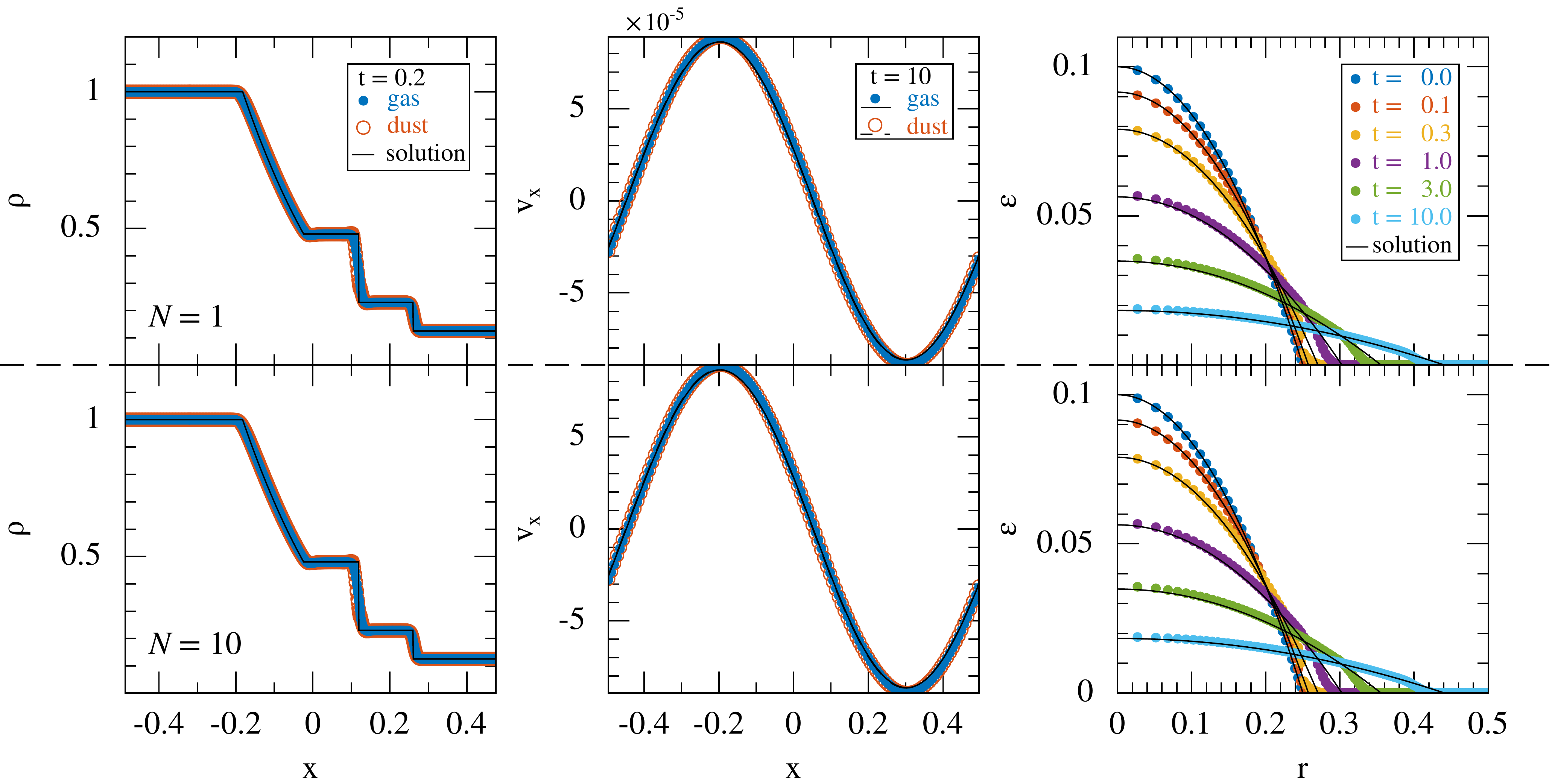}}
	\caption{Results from the \textsc{dustyshock} (left column), \textsc{dustywave} (middle column), and \textsc{dustydiffuse} (right column) tests as performed by \citetalias{Price/Laibe/2015} and \citet{Price/etal/2017}, but with our new array structure capable of handling multiple dust phases. The top row shows results when $N = 1$ while the bottom row contains simulations where the same dust phase has been split into $N = 10$ equal mass bins. As desired, the two cases are identical. Moreover, they match the results run prior to the \textsc{multigrain} implementation.}
	\label{fig:dust_tests}
\end{figure*}
By far, the most difficult part of implementing the \textsc{multigrain} method was expanding the relevant arrays in the code to accommodate the $N-1$ additional dust phases. To ensure that our new array structures cause no ill effects, we performed 3D versions of the \textsc{dustyshock}, \textsc{dustywave}, and \textsc{dustydiffuse} tests from \citetalias{Price/Laibe/2015}. Interested readers can find the setup details for these tests in \citet{Price/etal/2017}. The results from these tests are shown in the top row of \cref{fig:dust_tests}. Importantly, we found that the results calculated with and without our new array structure matched to within machine precision. Note that this agreement shows that \cref{eq:sph_density,eq:sph_momentum,eq:sph_dustfrac,eq:sph_energy} reduce numerically to the $N=1$ case, analogous to what we observed with the continuum equations. 

We then added an additional layer of complexity by splitting the single dust phase in each of the above tests into $N$ equal bins and evolving them as if they were $N$ different dust phases. This new setup can be achieved by setting $\epsj = \epsilon/N$ and $K_j = K$. Separating the fluid into mass bins does not alter the physics of the problem, just the numerical method by which it is modelled. Therefore, we should recover the $N=1$ solution (modulo numerical errors from calculating and combining quantities differently). \cref{fig:dust_tests} (bottom row) shows the results from the \textsc{multigrain} calculations. Again we found that the \textsc{dustyshock}, \textsc{dustywave}, and \textsc{dustydiffuse} tests agreed with the $N=1$ cases to within machine precision.

\subsection{Testing the general case}

It seems like the next logical test would be to extend one or more of the tests above to the general case of $N$ different dust phases. However, there is a fundamental difference in the way the drag is calculated for these tests and the way we have assumed the drag will be calculated when using the equations derived in this paper. Whereas the tests above use a constant drag coefficient $K$ for the entire fluid, the equations in \cref{sec:diffusion_approximation,sec:SPH_formulation} are optimised for physical dust grains in the Epstein drag regime where the equivalent drag constant \labelcref{eq:drag_constant_j} changes with grain size. We could reformulate the tests and their solutions to accept a unique value of $K$ for each dust phase, but this would require altering \cref{eq:sph_dustfrac,eq:sph_energy} --- the very equations we are trying to verify. Therefore, for the general case, we need a test requiring physical grain sizes and drag coefficients.

\subsection{Dust settling in a protoplanetary disc}
\label{sec:dusty_settle}

The dust settling test from \citepalias{Price/Laibe/2015} is an ideal candidate for testing the general case because it mimics one of the environments the \textsc{multigrain} method is designed to simulate, namely the settling of small dust grains in protoplanetary discs.

\subsubsection{Initial conditions}
\label{sec:dusty_settle_setup}

We simulate a disc-like environment at a radius $r = 50\,\mathrm{au}$ using a thin, vertical (Cartesian) column of gas in near-hydrostatic equilibrium with an external acceleration in the form of
\begin{equation}
	\mathbf{a}_\mathrm{ext} = - \frac{\mathcal{G} M z}{\left(r^2+z^2\right)^{3/2}}\mathbf{\hat{z}},
	\label{eq:vertical_gravity}
\end{equation}
where $\mathcal{G}$ is Newton's gravitational constant, $M$ is the stellar mass, and $z$ is the `vertical' coordinate along the length of the column ($x$ and $y$ represent the two shorter dimensions of the column). The gas density of the column is given by
\begin{equation}
	\rho_{\mathrm{g}}(z) = \rhogO \, \mathrm{exp} \left[-\frac{z^2}{2H^2}\right],
	\label{eq:isothermal_density_profile}
\end{equation}
where we choose $H/r = 0.05$, giving a disc scale height of $H = 2.5\,\mathrm{au}$. We assume an isothermal equation of state with $P = \cs^2 \rhog$, where $\cs \equiv H \Omega$ and $\Omega \equiv \sqrt{\mathcal{G} M / r^3}$, corresponding to an orbital time $t_\mathrm{orb} \equiv 2\pi/\Omega \approx 353\,\mathrm{yrs}$. We adopt code units with a distance unit of $10\,\mathrm{au}$, mass in solar masses and time units such that $\mathcal{G} = 1$. These choices give an orbital time of $\approx 70.2$ in code units. 

The particles are initially placed on a close-packed lattice using $100 \times 86 \times 78 = 670\,800$ particles in the domain $[x,y,z] \in [\pm1,\pm0.75,\pm0.65]$. We then stretch the particles in $z$ using the method described in \citet{Price/2004} to give the density profile given in \cref{eq:isothermal_density_profile}. We set $\rho_{\mathrm{g},0}$ to $10^{-3}$ in code units ($\approx 6 \times 10^{-13}\,\mathrm{g/cm}^3$ in physical units), corresponding to a particle mass in code units of $2.42 \times 10^{-9}$. We use periodic boundary conditions in all directions, but set the boundary in $z$ at $\pm 10 H$ in order to avoid periodicity in the vertical direction.

\begin{figure*}
	\centering{\includegraphics[width=\textwidth]{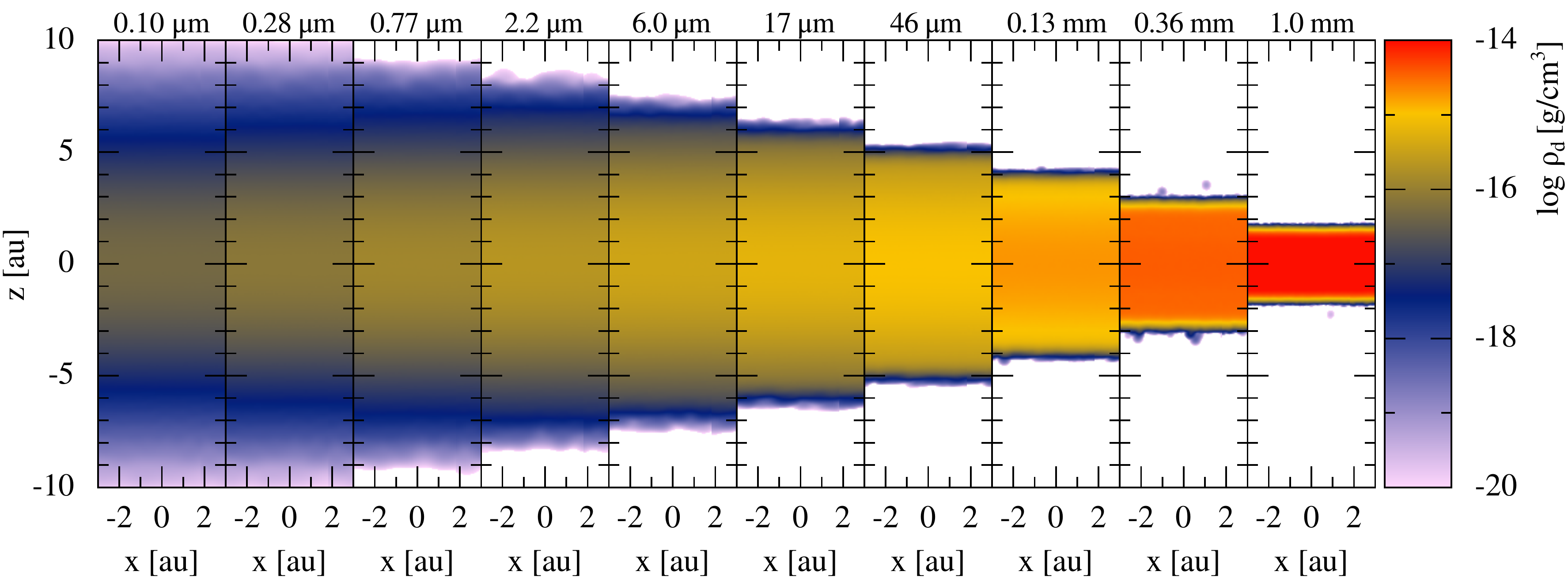}}
	\caption{Ten dust densities from a \textsc{multigrain} simulation after having settled for 15 orbits in a 3D vertical column of a protoplanetary disc at $r = 50\,\mathrm{au}$ (assuming H/r = 0.05; so H = 2.5 au) using $100 \times 86 \times 78	 = 670\,800$ simulation particles. The grain size and initial dust fraction for each phase is listed in \cref{tab:initial_vals}. Large dust grains efficiently settle towards the disc mid-plane, but still have a much lower density than the smaller dust grains because the global number density of the larger grains is lower. Our \textsc{multigrain} simulation is $\sim5\times$ faster to run than 10 single-phase simulations run serially (see \cref{sec:performance}).}
	\label{fig:settling_test}
\end{figure*}
\begin{figure*}
	\centering{\includegraphics[width=\textwidth]{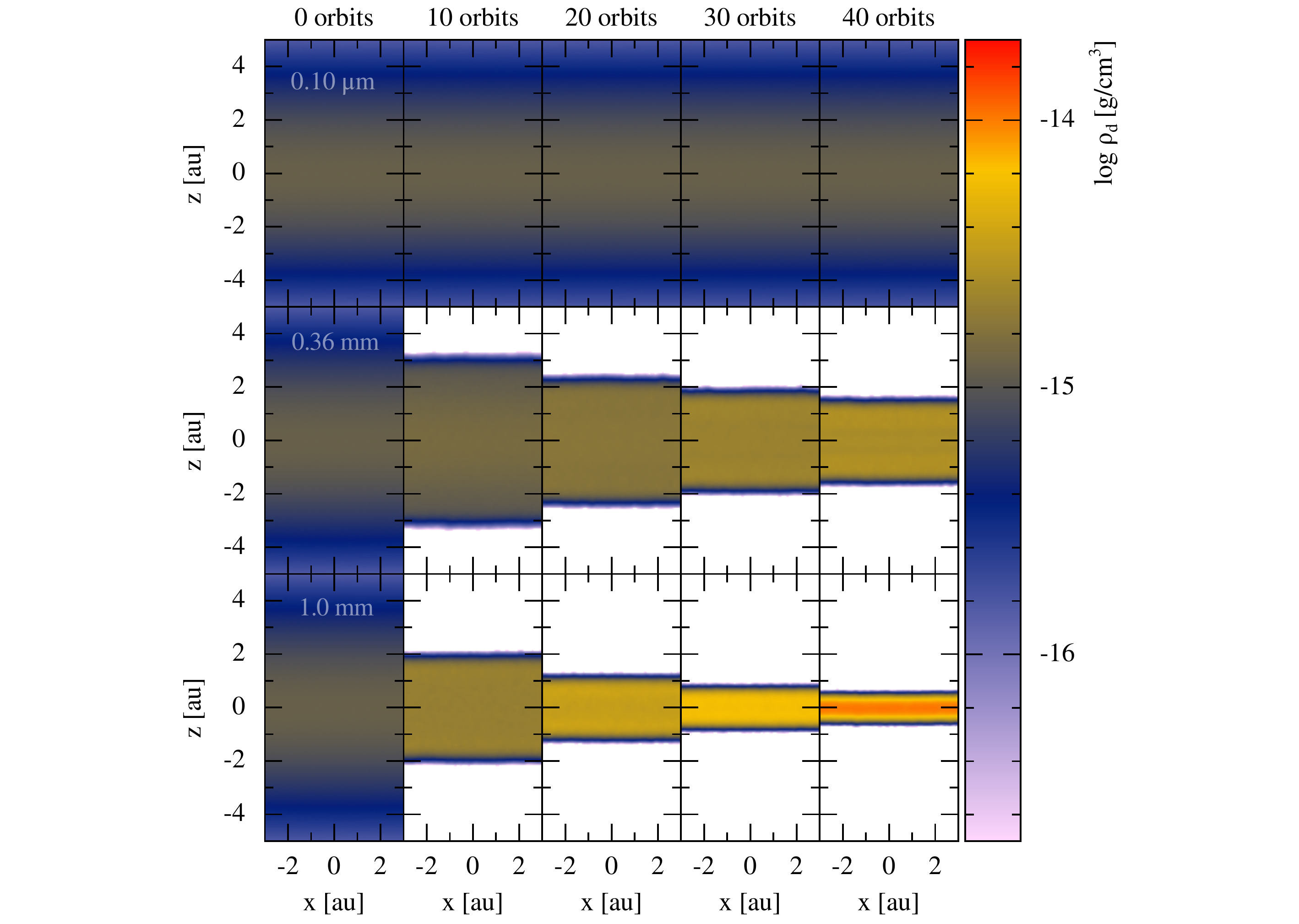}}
	\caption{Time evolution of the the densities of three dust phases ($j = [1,9,10]$). The initial conditions in this simulation were the same as in \cref{fig:settling_test}, except with equal dust fractions ($\epsj = \epsilon/N$) to make the relative density enhancement within and between dust phases more visible. We have also adjusted the colourbar in order to allow direct comparison with the settling tests performed by \citetalias{Price/Laibe/2015} and \citet{Price/etal/2017}. Note that the density enhancement due to settling has a shallower dynamic range than the built-in density gamut created by our grains-size distribution (see \cref{fig:settling_test}).}
	\label{fig:compare_PL15}
\end{figure*}
\begin{figure*}
    \centering
    \begin{minipage}[b]{\columnwidth}
        \centering
        \includegraphics[width=\columnwidth]{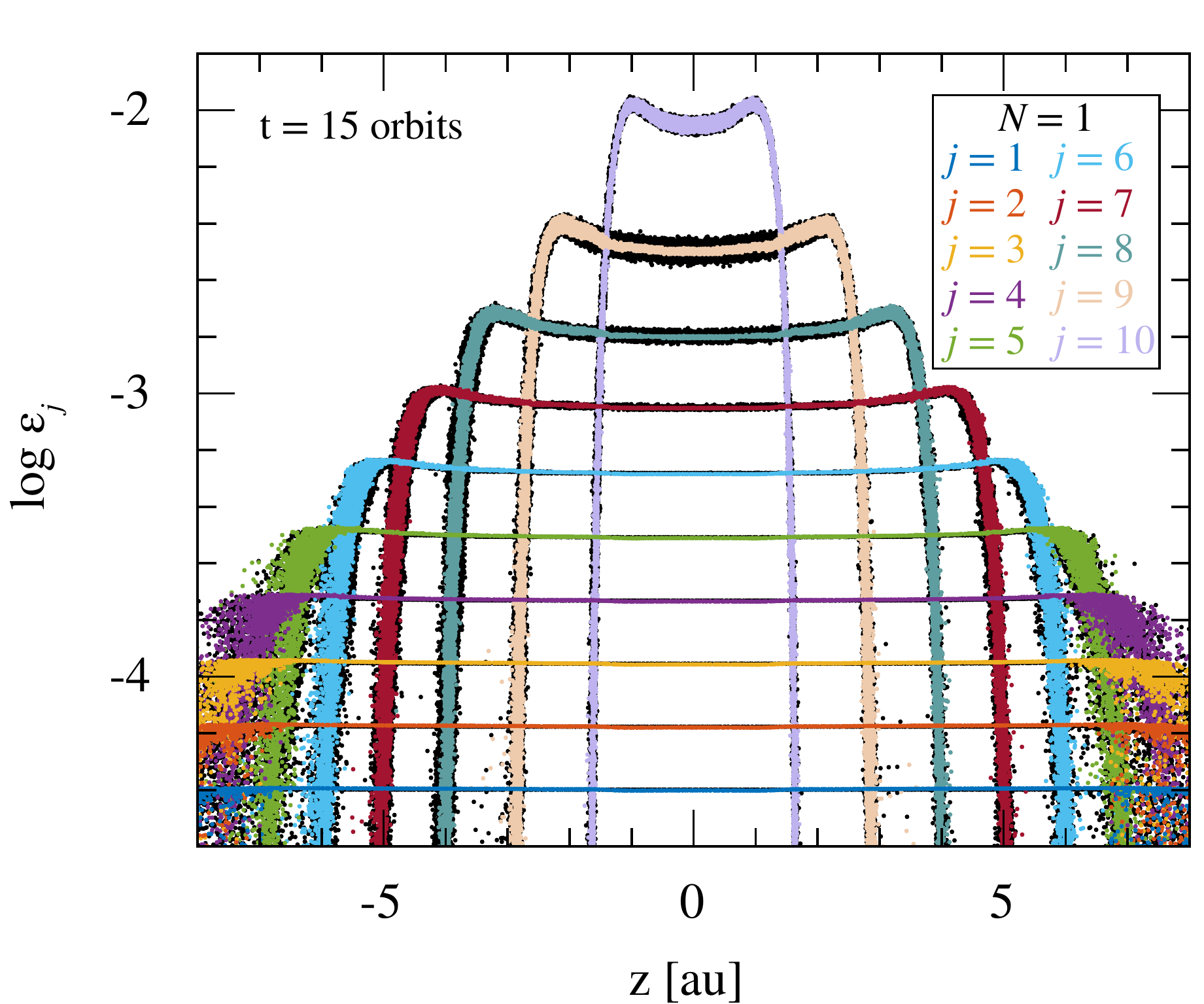}
    \end{minipage}%
    \hfill%
    \begin{minipage}[b]{\columnwidth}
    \centering
        \centering
        \includegraphics[width=\columnwidth]{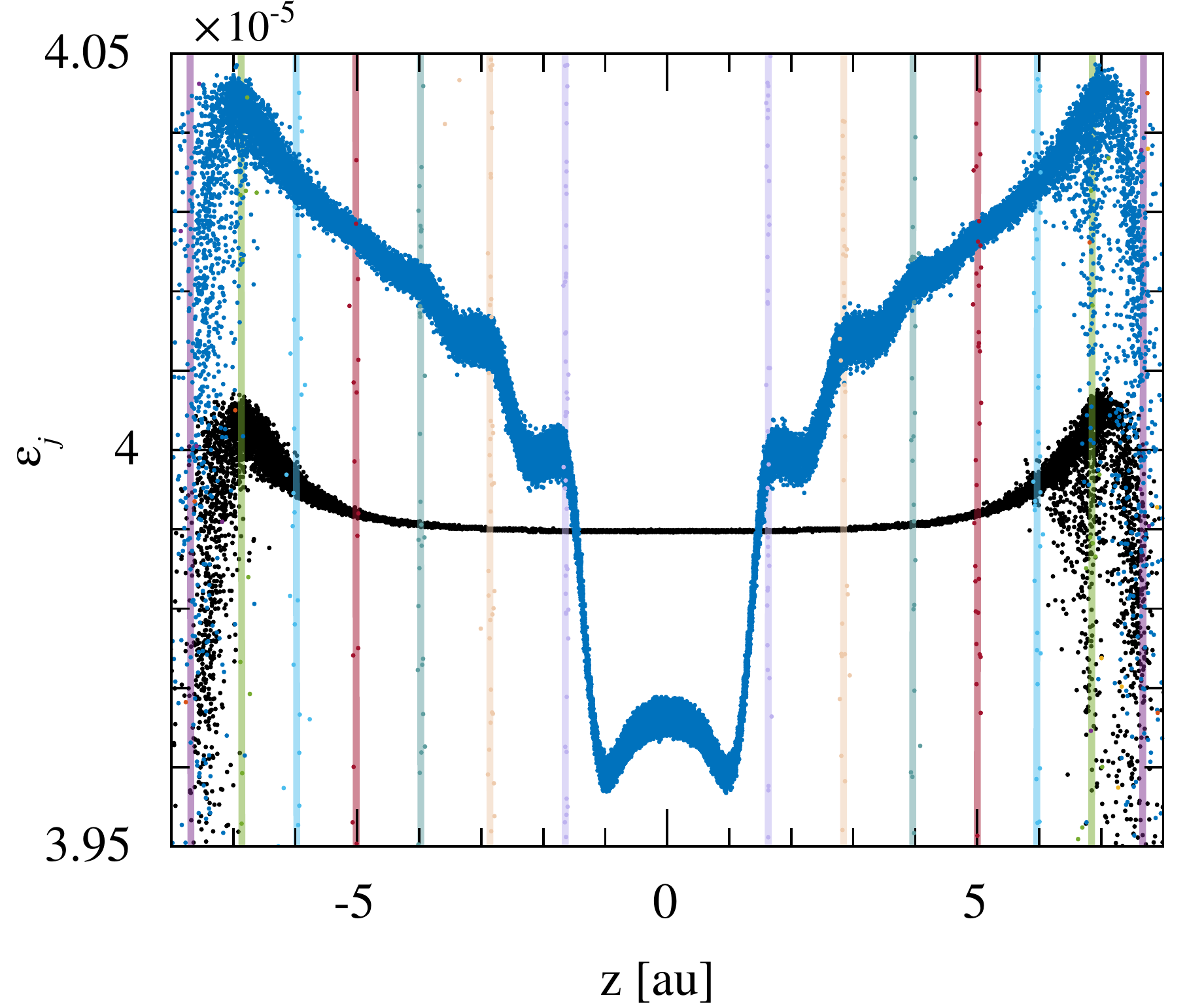}
    \end{minipage}\\[-7pt]
    \begin{minipage}[t]{\columnwidth}
        \caption{Comparison of dust fractions after 15 orbits when calculated by 10 single-phase simulations (black points) vs 1 \textsc{multigrain} simulation (coloured points). Not only does the \textsc{multigrain} method recover the correct solution, but the dispersion in $\epsj$ is equal to or better than the single-phase simulations.}
        \label{fig:back_reaction}
    \end{minipage}%
    \hfill%
    \begin{minipage}[t]{\columnwidth}
        \caption{A zoom in of the $s=0.1\,\mu\mathrm{m}$ grains in \cref{fig:back_reaction}, highlighting the non-linear coupling between dust phases captured in a \textsc{multigrain} simulation (blue points) compared to the single-phase simulation (black points). The location of the peaks (resp. troughs) seen in $\epsilon_1$ correlate with the outer edges (resp. density peaks) of the other phases. We have added semi-transparent lines to help identify the location of the other phases in the figure. With the exception of the largest grain size, the remaining phases exhibit similar discrepancies with their single-phase counterpart.}
        \label{fig:zoomed-in}
    \end{minipage}%
\end{figure*}

We relaxed the density profile by running the code for 15 orbits with artificial viscosity, at which point we added $N = 10$ distinct dust phases to the system. We created a cell-edge, logarithmic grid from $\smin$ to $\smax$ with grid cells of width $\Delta \log s = \frac{1}{N} \log_{10}\left( \smax/\smin \right)$. Then we assigned $\sj$ by taking the square root of the product of the cell's endpoints --- thereby skewing the `typical' grain size for each cell towards the smaller, more numerous dust grains. Each dust phase was distributed throughout the disc with an initially uniform dust fraction. We constrained the total dust fraction to be $\epsilon = 1/101$ (corresponding to a dust-to-gas ratio of 0.01) and set the magnitudes of $\epsj$ according to the differential power-law distribution
\begin{equation}
	\mathrm{d} \epsilon = \epsilon_0 s^{3-p} \mathrm{d} s, \qquad \mathrm{for} \quad \smin \leq s \leq \smax,
	\label{eq:MRN_distribution}
\end{equation}	
where $\mathrm{d} \epsilon$ is the differential dust fraction with respect to grain size, $\epsilon_0$ is a normalisation factor, and $p$ is the usual power-law index for number density as a function of grain size \citep[e.g.][]{Mathis/Rumpl/Nordsieck/1977}. In particular, $\epsj$ is determined by integrating \cref{eq:MRN_distribution} across each grain-size cell and then normalising their combined sum via \cref{eq:def_epsilon}. Assuming $p = 3.5$, we set $\smin \approx 0.0599\,\mu\mathrm{m}$ and $\smax \approx 1.67\,\mathrm{mm}$ such that the smallest simulated grain size is $0.1\,\mu\mathrm{m}$ and the largest simulated grain size is $1\,\mathrm{mm}$. The initial values for $\sj$ and $\epsj$ in this test are listed in \cref{tab:initial_vals}.
\begin{table}
	\centering
	\caption{The initial values for $\sj$ and $\epsj$ used in the settling test assuming a power-law distribution in grain sizes ranging from $\smin = 0.1\,\mu\mathrm{m}$ to $\smax = 1\,\mathrm{mm}$ with a power-law index of $p=3.5$.}
	\label{tab:initial_vals}
	\sisetup{table-format = 1.2,table-auto-round = true}
	\begin{tabular*}{0.75\columnwidth}
		{@{\extracolsep{\stretch{1}}}
			S[table-format=2.0]
			S[table-format=1.2e-1,scientific-notation=true]
			S[table-format=1.2e-2,scientific-notation=true]
		@{}} \toprule
		{$j$} 	&	{$\sj\,$[cm]}			&	{$\epsj$}				\\\midrule
		1	&	1.000000000000000E-5	&	3.989418407119701E-5	\\
		2	&	2.782559402207126E-5	&	6.654750988032161E-5	\\
		3	&	7.742636826811278E-5	&	1.110079369806909E-4	\\
		4	&	2.154434690031882E-4	&	1.851723993109608E-4	\\
		5	&	5.994842503189409E-4	&	3.088861787652376E-4	\\
		6	&	1.668100537200059E-3	&	5.152532007319657E-4	\\
		7	&	4.641588833612777E-3	&	8.594941409350411E-4	\\
		8	&	1.291549665014883E-2	&	1.433722638214047E-3	\\
		9	&	3.593813663804626E-2	&	2.391593503000737E-3	\\
		10	&	0.100000000000000E-0	&	3.989418407119701E-3	\\\bottomrule
	\end{tabular*}
\end{table}

\subsubsection{Results}
\label{sec:settling_test_results}

\begin{figure*}
    \centering
    \begin{minipage}[b]{\columnwidth}
        \centering
        \includegraphics[width=\columnwidth]{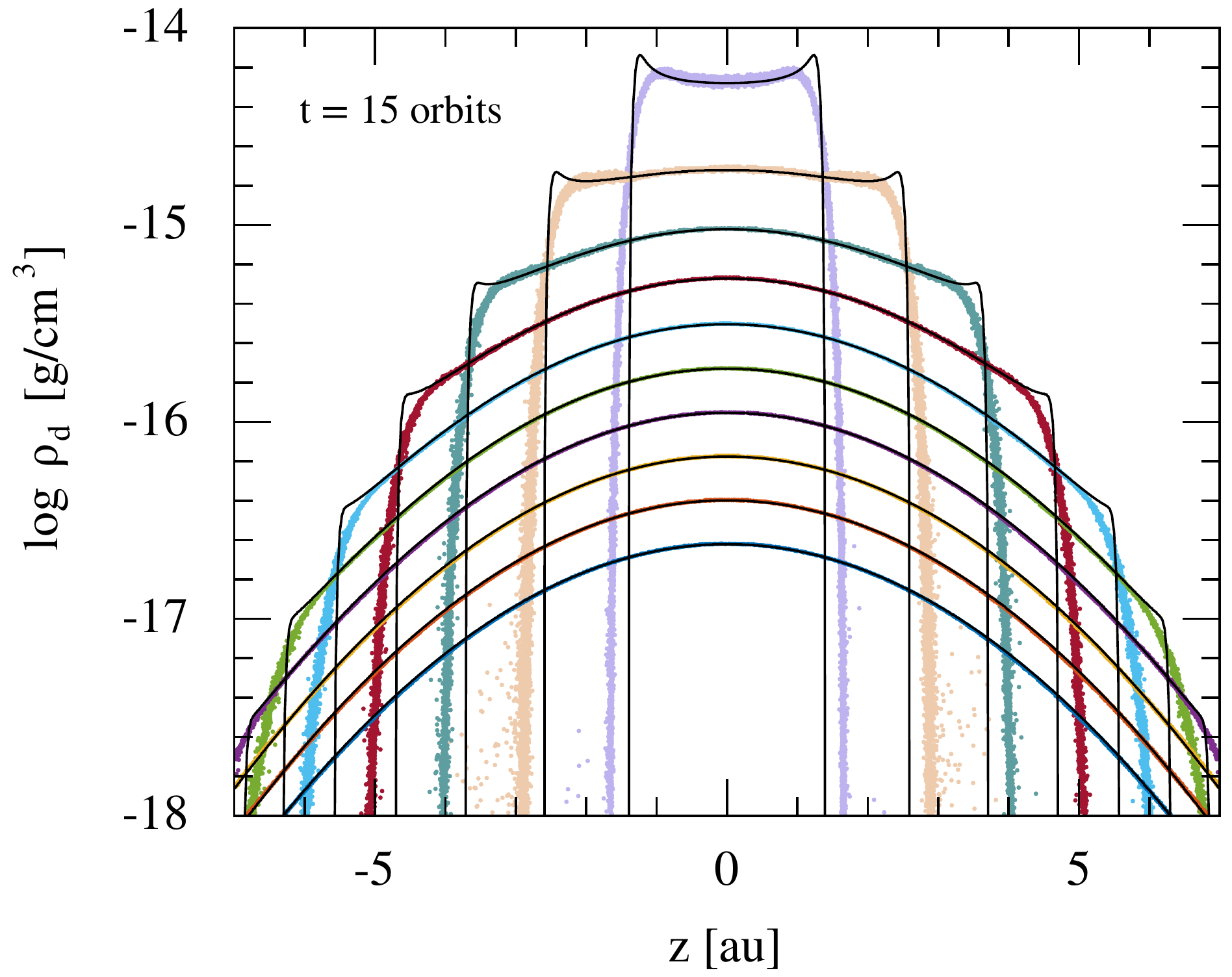}        
    \end{minipage}%
    \hfill%
    \begin{minipage}[b]{\columnwidth}
    \centering
        \centering
        \includegraphics[width=\columnwidth]{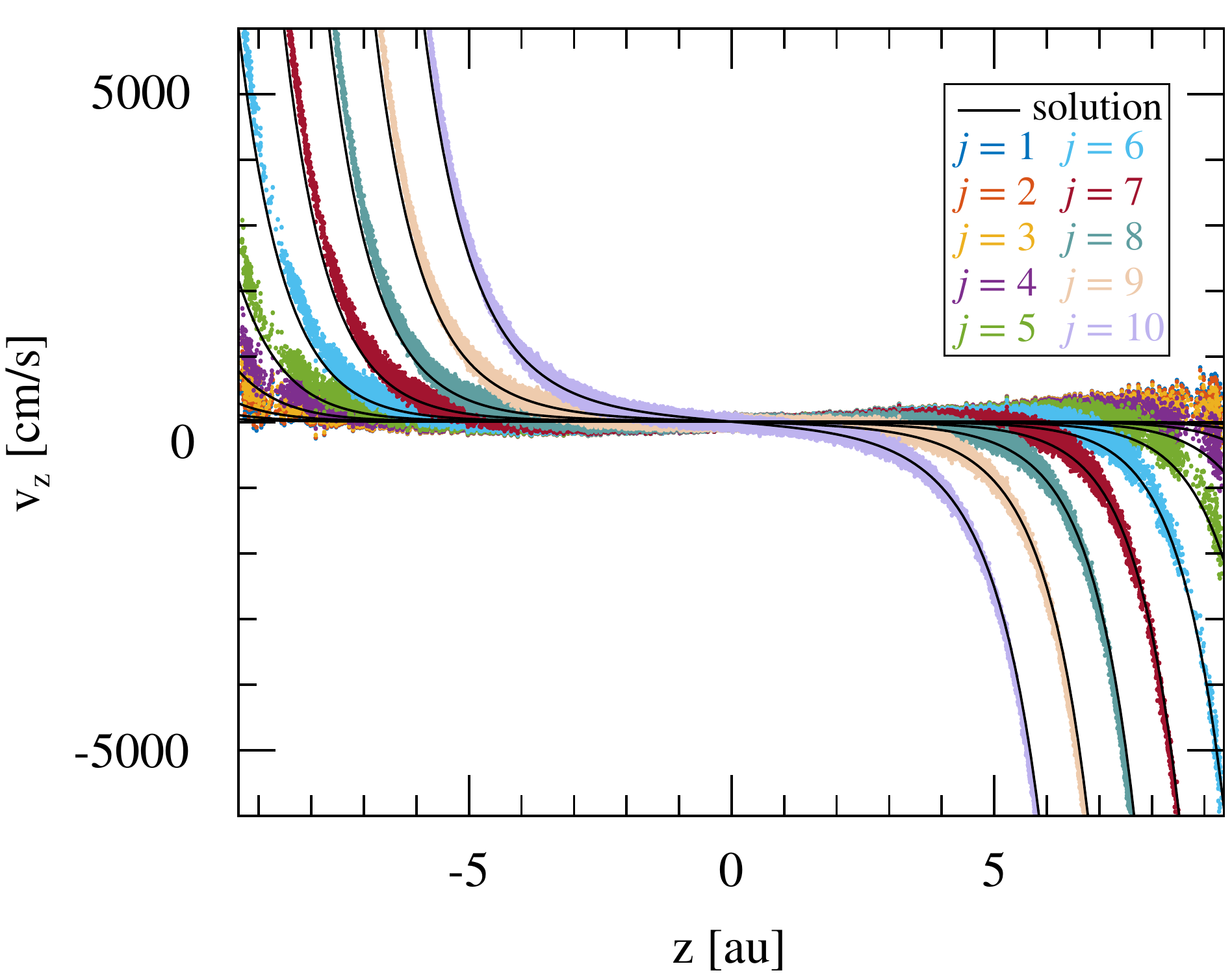}
    \end{minipage}\\[-7pt]
    \begin{minipage}[t]{\columnwidth}
        \caption{Dust densities for dust settling test in \cref{fig:back_reaction}. The coloured points are from our \textsc{multigrain} simulation while the black solution curves come from solving the \cref{eq:rhod_nodim,eq:vd_nodim} on a grid. The over-smoothing of the dust fronts is an artefact of tracking the total mass rather than just the dust mass. However, the location of the dust fronts and the enhancement of the mid-plane densities are well captured by the \textsc{multigrain} method.}
        \label{fig:density_curves}
    \end{minipage}%
    \hfill%
    \begin{minipage}[t]{\columnwidth}
        \caption{Settling velocities for the dust settling test in \cref{fig:back_reaction}. Again, the coloured points are results from our \textsc{multigrain} calculations, but the black curves are now the single-phase analytic solutions from \cref{eq:vd_solution}. Slight discrepancies are visible due to the coupling between dust phases and the fact that the gas is not stationary, both of which are ignored in the analytic solution.}
        \label{fig:velocity_curves}
    \end{minipage}%
\end{figure*}

After adding the dust, we ran the simulation for an additional 15 orbits. The resulting dust density for each of the different phases is shown in \cref{fig:settling_test}. As expected, the settling efficiency is proportional to the size of the dust grains, thus enhancing the mid-plane density of the larger grains. However, visually separating this density enhancement is difficult in \cref{fig:settling_test} because the initial density distribution increases by a factor of 100 from $\smin$ to $\smax$. Although the continuum density distribution is a \textit{de}creasing function with respect to grain size ($\propto s^{3-p} = 1/\sqrt{s}$), integrating over each cell to include the mass from non-simulated grains steepens the power-law by an additional power of $s$ such that $\rhodj \propto \sqrt{s}$.

In order to better show how the mid-plane density is affected by settling, we setup and ran a second simulation where the dust fractions were all equal, i.e. $\epsj = \epsilon/N$. \Cref{fig:compare_PL15} shows the resulting time evolution of the dust density for phases $j = [1,9,10]$. This time we clearly see that settling increases the dust density relative to its initial state and at a rate that is commensurate with its settling efficiency. These results are in good agreement with previous settling tests performed in the literature \citep[\citetalias{Price/Laibe/2015};][]{Hutchison/etal/2016,Price/etal/2017}, albeit with a smaller initial dust fraction.

As a further benchmarking exercise, we ran 10 single-phase simulations using the initial conditions from \cref{tab:initial_vals} and compared them to the results from the multiphase test above (see \cref{fig:back_reaction}). Although the two scenarios are not strictly equivalent --- the single-phase simulations do not include the backreaction from the $N$-$1$ other phases --- the global solutions still match because (i) the majority of the disc mass resides in the gas and (ii) the gas is essentially motionless throughout the simulation (see, e.g., the top row in \cref{fig:compare_PL15}, which can be used as a proxy for the gas). In the limit of zero backreaction and a stationary gas phase, the system can be modelled analytically and numerically using a simplified set of fluid equations. \cref{sec:settling_test_derivation} gives the full analysis.

The large-scale agreement we see in \cref{fig:back_reaction} does not extend to smaller scales. In \cref{fig:zoomed-in} we zoom in on the $s=0.1\,\mu\mathrm{m}$ grains to illustrate the substructure in $\epsj$ that develops as a result of the backreaction included from other dust phases. These differences between the single-phase and \textsc{multigrain} simulations continue to grow with time. Therefore, single-phase simulations should be used with caution in situations involving turbulent gas dynamics and/or long timescales.

The differences we observe in \cref{fig:zoomed-in} are small, but prevent the test from being truly rigorous. One way of making the indirect coupling between dust phases vanishingly small is to concentrate all of the dust mass into the smallest grains which remain fixed to the gas, i.e. stationary. We found that by using a power-law index of $p = 6.5$, we could concentrate $\sim 99\,\%$ of the dust mass in the two smallest phases (with $j = 1$ accounting for $\sim 92\,\%$). Under these new conditions, our single-phase and \textsc{multigrain} simulations were a near perfect match at all scales. The only visible difference between the two scenarios was a minor reduction in dispersion in some of the \textsc{multigrain} phases, similar in magnitude to what is seen in \cref{fig:back_reaction}.

As an interesting aside, the steeper power-law index of $p = 6.5$ produces a 10 order-of-magnitude gap between the mid-plane densities of the largest and smallest dust grains. Happily, roundoff errors do not appear to corrupt the results in this situation, which we attribute to the fact that each $\epsj$ is evolved separately. While the dust fractions are combined to calculate the gas properties, the gas-dust interaction depends only on the ratio of their masses. That is, the gas is not sensitive to tiny fluctuations in $\epsilon$ that may be introduced by loss of precision when combining $\epsj$ of very different magnitudes. 

So far we have relied on comparing our \textsc{multigrain} results with single-phase simulations. In \cref{fig:density_curves}, we return to using the initial conditions from \cref{tab:initial_vals} and compare our \textsc{multigrain} solution to a grid-based numerical solution described in \cref{sec:settling_numeric}. The settling fronts in our SPH simulations match the simplified solutions to better than a few percent for all except the largest grain sizes (see \cref{tab:errors}). As pointed out by \citetalias{Price/Laibe/2015}, the resolution follows the total mass rather than the dust mass, so it tends to over smooth the density peaks in the dust. Despite the smoothing, the locations of the settling fronts and the densities within the disc match very well. The L$_2$ errors scale with the grain size (see column 2 in \cref{tab:errors}) and are a result of the over-smoothed dust peaks and the increased dispersion in the density at larger grain sizes.

In \cref{fig:velocity_curves} we compare our \textsc{multigrain} simulation to the analytic solution in \cref{eq:vd_solution}. While we find L$_2$ errors of order 0.1--1\% for grain sizes $> 10\,\mu$m, there is a steady decline in accuracy as grain size decreases (see column 3 in \cref{tab:errors}). This progressive departure from the analytic model is a reflection of the fact that the gas is not completely stationary. Fluctuations in the gas velocity create a size-dependent velocity dispersion in the dust that primarily affects the smaller grain sizes. The larger dust grains, that are less susceptible to these fluctuations in the gas, exhibit less dispersion and better agreement with the analytic solution.

\subsubsection{Performance}
\label{sec:performance}
We ran each of the test simulations above using OpenMP on eight cores from a single node. We found that our \textsc{multigrain} simulations with $N=10$ dust phases were a factor of two slower than one single-phase simulation, thus making the \textsc{multigrain} simulations five times faster than their single-phase equivalent. This scaling improves as $N$ increases, provided there is enough memory to handle the large array sizes. For example, we found the \textsc{multigrain} method to be $\approx 13$ times faster when $N=100$. We expect even better performance ratios relative to multi-fluid simulations because multi-fluid methods require $N$ times more simulation particles and often an added overhead for implicit timestepping \citepalias[explicit multi-fluid methods are impossibly slow for most of the grain sizes considered in this study; see][]{Price/Laibe/2015}. Finally, because the \textsc{multigrain} method reuses the same simulation particles for all $N$ dust phases, it requires less post processing and, when $N=10$, uses 55 per cent less disk space than an equivalent set of single-phase simulations (65 per cent less when $N=100$). Files in which $\deltavj$ is not written to disk,\footnote{In the diffusion approximation, $\deltavj$ is a calculated quantity needed for recovering the gas and dust velocities during post processing. However, as $\deltavj$ is not required in any of the evolution equations, we often omit writing it to disk in order to save space.} are reduced by an additional 15 per cent.
\begin{table}
	\centering
	\caption{$L_2$ errors, computed by \textsc{splash} \citep{Price/2007}, between the \textsc{multigrain} dusty settling test from \cref{sec:dusty_settle} and the analytic/numerical solutions from \cref{sec:settling_test_derivation}. Column 2 is obtained using the numeric dust densities from \cref{sec:settling_numeric} and column 3 using the analytic dust velocities from \cref{eq:vd_solution}.  Large $L_2$ errors are caused mainly by inconsistencies between the analytical and numerical models. That our density errors are lowest where our velocity errors are largest (and vice versa) indicates that our \textsc{multigrain} solution is valid.}
	\label{tab:errors}
	\sisetup{table-format = 1.2,table-auto-round = true}
	\begin{tabular*}{0.85\columnwidth}
		{@{\extracolsep{\stretch{1}}}
			S[table-format=1.2e-1,scientific-notation=true]
			S[table-format=1.2e-1,scientific-notation=true]
			S[table-format=1.2e-2,scientific-notation=true]
		@{}} \toprule
		{$s$ [cm]} 	&	{$L_2$ density errors}	&	{$L_2$ velocity errors}	\\\midrule
		1.000000000000000E-5	&	4.92880E-3	&	1.46553E-0	\\
		2.782559402207126E-5	&	4.89849E-3	&	5.26670E-1	\\
		7.742636826811278E-5	&	4.81021E-3	&	1.89265E-1	\\
		2.154434690031882E-4	&	5.09736E-3	&	6.80182E-2	\\
		5.994842503189409E-4	&	5.05506E-3	&	2.44719E-2	\\
		1.668100537200059E-3	&	6.44138E-2	&	8.89909E-3	\\
		4.641588833612777E-3	&	1.24372E-2	&	3.50069E-3	\\
		1.291549665014883E-2	&	3.43672E-2	&	1.91667E-3	\\
		3.593813663804626E-2	&	7.89282E-2	&	1.60872E-3	\\
		0.100000000000000E-0	&	9.42728E-2	&	1.56667E-3	\\\bottomrule
	\end{tabular*}
\end{table}

\subsection{Radial drift in a protoplanetary disc}
\label{sec:radial_drift}

The dusty settling test in the previous section remains well approximated by single-phase methods. To demonstrate that our algorithm also works in regimes of strong backreaction, we computed the radial drift velocities for two dust phases in a protoplanetary disc with conditions such that the inward migration of the larger grains induces a discernable outward migration of the smaller grains.

\subsubsection{Analytic solution}
\label{sec:analytic_sol_radial_drift}

An analytic solution for multiple dust phases migrating in an inviscid disc was derived by \citet{Bai/Stone/2010b}. Neglecting vertical gravity, they show that the hydrostatic equilibrium equations can be written in block matrix form as follows:
\begin{gather}
	\begin{pmatrix} 
		\matrixbf{I} + \matrixbf{\Gamma} & -2 \matrixbf{\Lambda}  
	\\ 
		\matrixbf{\Lambda}/2 &  \matrixbf{I} + \matrixbf{\Gamma}
	\end{pmatrix}
	\begin{pmatrix}
		\matrixbf{\mathcal{V}}_r
	\\
		\matrixbf{\mathcal{V}}_\phi
	\end{pmatrix}
 	=
	- \eta v_K
  	\begin{pmatrix}
		\matrixbf{0}
	\\
		\matrixbf{1}
	\end{pmatrix},
	\label{eq:eom_radial_drift}
\end{gather}
where $\matrixbf{I}$ is the identity matrix, $\matrixbf{\mathcal{V}}_r \equiv \left( v_{1r},v_{2r},\ldots,v_{nr} \right)^\intercal$ and $\matrixbf{\mathcal{V}}_\phi \equiv \left( v_{1\phi},v_{2\phi},\ldots,v_{n\phi} \right)^\intercal$ are the radial and azimuthal velocities for each dust phase, respectively. The matrix $\matrixbf{\Lambda} \equiv \text{diag}\left\{\text{St}_1,\text{St}_2,\ldots,\text{St}_n \right\}$ is a diagonal matrix of the Stokes numbers for \textit{uncoupled} dust phases (i.e. $\St = \tsone \OmegaK$), while $\matrixbf{\Gamma} \equiv \left(\matrixbf{\mathcal{E}},\matrixbf{\mathcal{E}},\ldots,\matrixbf{\mathcal{E}} \right)^\intercal$ is a matrix made up of the dust-to-gas ratios, where $\matrixbf{\mathcal{E}} \equiv \left(\mathcal{E}_1,\mathcal{E}_2,\ldots,\mathcal{E}_n \right)^\intercal$ and $\mathcal{E}_j \equiv \rhodj/\rhog = \epsj/(1-\epsilon)$. \citet{Bai/Stone/2010b} provide a closed-form solution to \cref{eq:eom_radial_drift}; however, we found it more convenient to solve it numerically.

\subsubsection{Setup}
\label{sec:radial_drift_setup}

We setup a 3D, locally isothermal gas disc using the following power-law parameterisations \citep[see, e.g.,][]{Laibe/Gonzalez/Maddison/2012}
\begin{align}
	& c_{\text{s}}(r) = c_{\text{s},1\text{au}} \left( \frac{r}{1\,\text{au}} \right)^{-q/2},
\\
	& H_\text{g}(r) = H_{\text{g},1\text{au}} \left( \frac{r}{1\,\text{au}} \right)^{3/2- q/2},
\\
	& \Sigma_\text{g}(r) = \Sigma_{\text{g},1\text{au}} \left( \frac{r}{1\,\text{au}} \right)^{-p},
\\
	& \rho_\text{g}(r,z) = \frac{\Sigma_\text{g}}{\sqrt{2\pi}H} \exp{\left[ -\frac{z^2}{2 H^2}\right]},
\end{align}
where $\Sigma_\text{g}$ is the local surface density for the gas, quantities with the subscript `$1\,\text{au}$' are reference values measured at $r=1\,\text{au}$, and the parameters $p = 1$ and $q = 0.5$ are power-law exponents controlling the density and temperature (i.e. flaring) of the disc, respectively. We set the radial velocity to zero and correct the orbital velocities from pure Keplerian rotation to account for the radial pressure gradient in the disc,
\begin{equation}
	v_\phi = v_K(1 - \eta),
\end{equation}
where the pressure gradient parameter $\eta$ is given by
\begin{equation}
	\eta  \approx \frac{1}{4} \left(\frac{H_\text{g}}{r} \right)^2 \left[ 3 + 2 p+ q - \left( 3 - q \right) \left( \frac{z}{H_\text{g}} \right)^2  \right] ,
\end{equation}
to order $z^2/r^2$ \citep[see, e.g.,][]{Takeuchi/Lin/2002}.
We add the dust by assigning a uniform dust fraction to all of the particles. Because the diffusion approximation assumes the stopping time is much shorter than the dynamical timescale, we do not give the dust a separate azimuthal velocity.

The analytic solution from \citet{Bai/Stone/2010b} is 2D and assumes that dust resides in the mid-plane of the disc. As both gas density and gravity decrease with increasing $z$, we expect the dust at high altitudes to migrate slower than dust in the mid-plane. To compensate, we only compare migration rates for $|z| < H_\text{g}(r)$, we bin the particles radially into 50 logarithmically spaced bins (using the same binning method described for the grain-size distribution previously), and we average the radial velocities both azimuthally and vertically within each bin.

One final caveat remains: the steady-state analytic model assumes the disc is inviscid, whereas SPH disc simulations are inherently viscous. Normally we would relax our disc into a quasi-steady state and use our instantaneous gas and dust mid-plane densities as initial conditions for the model --- thereby allowing us to account for any non-steady-state processes like settling and/or migration. However, the lack of viscidity in the model produces rigid assumptions about the gas velocity that are not met in our viscous SPH simulations. As a result, we find that our simulation relaxes into a steady state that is substantially different to the analytic solution. To our knowledge, there is currently no analytic solution for radial velocities in viscous discs. Deriving such a solution goes beyond the scope of this paper; therefore, we will revisit the problem in a future study.

In the meantime, we can circumvent this incompatibility in the present study by using the initial state of the system, where we have full control over the velocities and we can mimic the conditions of an inviscid disc. Testing the initial conditions, albeit unorthodox, still yields valuable information about our method for two reasons. First, the terminal velocity approximation breaks down when the timestep is smaller than a few stopping times. Because the typical time for drift to relax is on the order of a few stopping times, we do not need to wait for the dust velocities to equilibrate. In other words, the full asymptotic radial velocities are obtained after the very first loop over the particles (what we call $t = 0$) when $P$, $\Tsj$, $\Ts$, $\deltavj$, and $\deltav$ are all calculated --- the quantities we use to construct the velocity profiles of the gas and dust. Secondly, the individual gas and dust properties are calculated (as opposed to being evolved). Therefore, the test is more sensitive to how we calculate the forces than how we evolve the mixture. Since our force prescription does not vary with time, the test is almost as useful at $t = 0$ as it would be once they system has reached a steady state.

\subsubsection{Results}
\label{sec:radial_drift_results}

\begin{figure*}
        \centering
        \includegraphics[width=\textwidth]{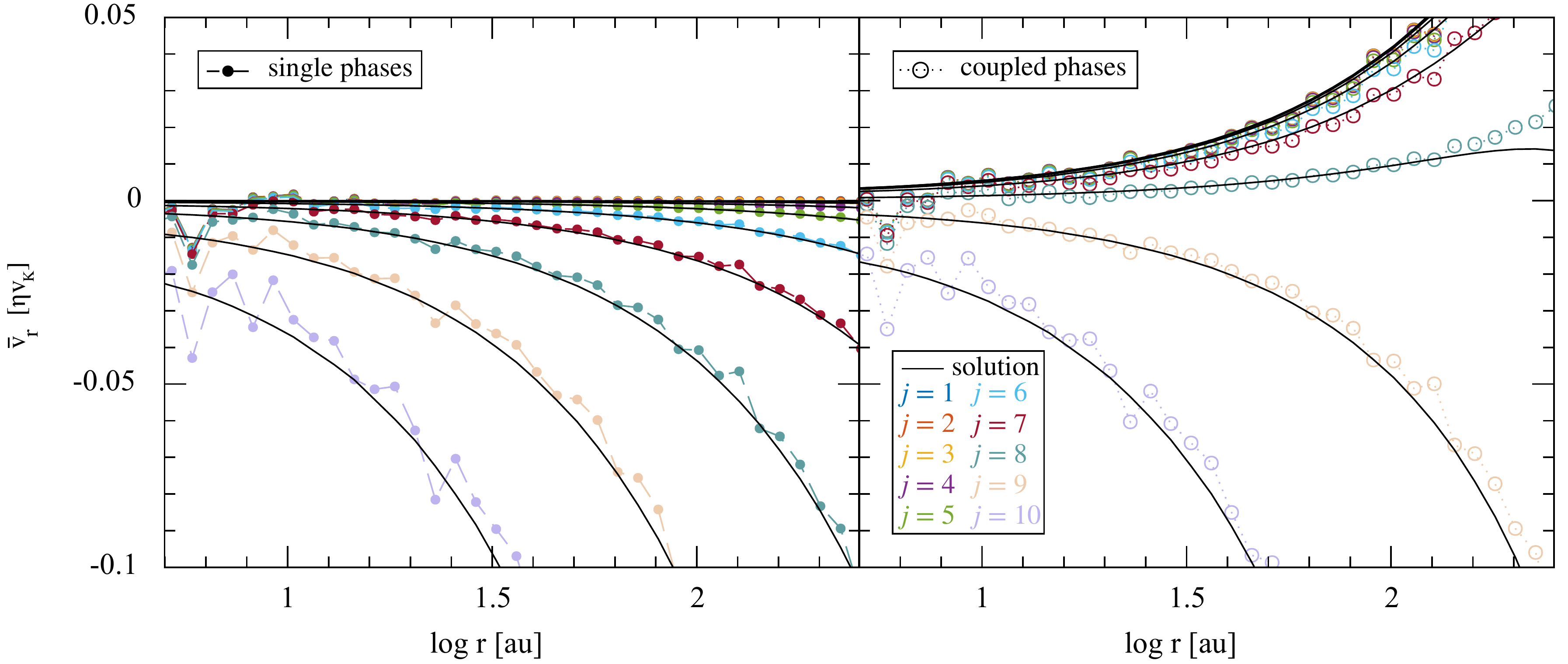}
        \caption{Mean radial drift velocities near the mid-plane of a protoplanetary disc for ten dust phases of different grain size, individually coupled with the gas (left) and simultaneously coupled (right). Coloured points (connected by either dashed or dotted lines for clarity) show the mean radial dust velocity calculated for each dust phase from the barycentric simulation data and solid black lines indicate the corresponding analytic solutions. The complete lack of outward migration in the left panel accentuates how single-phase simulations can miss dynamical effects caused by the presence of other dust phases. The right panel illustrates that the \textsc{multigrain} formalism is capable of resolving outward/inward migration velocities of different dust species within a single set of simulation particles.}
        \label{fig:radial_drift}
\end{figure*}

Using the same grain-size distribution from \cref{sec:dusty_settle_setup}, we setup a dusty protoplanetary disc around a solar mass star with an inner and outer radius of $r_{\text{in}} = 1\,\text{au}$ and $r_{\text{out}} = 300\,\text{au}$, respectively. The gas disc has the following reference values:  $c_{\text{s},1\text{au}} \approx 1.5 \, \text{km s}^{-1}$, $H_{\text{g},1\text{au}} = 0.05 \,\text{au}$, and $\Sigma_{\text{g},1\text{au}} \approx 166 \,\text{g cm}^{-2}$. The dust disc for each grain size is set equal to the gas in size and shape, but scaled in mass by the dust fractions listed in \cref{tab:dust_vals_migration}, such that, when all of the dust phases are included, the total dust mass comprises 1/3 of the total mass of the system (i.e. a dust-to-gas ratio of $\mathcal{E} = 0.5$). For the ten single-phase calculations, it is not possible to simultaneously match the dust fractions, dust-to-gas ratios, and the gas/dust densities of the \textsc{multigrain} case. Therefore, we chose to keep the respective dust fraction and the total surface density of the disc the same, while allowing the dust-to-gas ratio for each dust phase to change as needed. This discrepancy with the 
\textsc{multigrain} calculations results in different surface densities for the gas and dust, but the effects are unimportant in this context since outward migration of dust in a single-phase simulation can only be achieved by drastically changing the structure of the disc (e.g. with a radially increasing pressure profile).

The left and right panels in \cref{fig:radial_drift} show the mean radial dust velocities for the individual and combined cases, respectively. Coloured points are the velocities calculated from the SPH mixture while the solid black lines show the corresponding analytic solutions. Although the two largest grain sizes exhibit only minor changes to their velocities after the addition of the other dust species, the eight smaller sizes experience a complete reversal in migration direction. This change in sign is caused by the exchange of angular momentum as the larger grains drag the sub-Keplerian gas into faster orbits, thereby pushing the gas radially outward. The smaller dust grains, who are more sensitive to changes in the gas flow, are then carried outward along with the gas. Outward migration of dust in a disc with a radially decreasing pressure gradient cannot be replicated with only one dust phase; conservation of angular momentum requires one or more phases to radially contract as the others expand. Importantly, the \textsc{multigrain} formalism correctly resolves the velocities for both outward and inward migrating species.

The relative angular velocity between the gas and dust varies with height. In fact, $\eta$ changes sign at $z \approx 1.5 H_\text{g}$, meaning that dust particles rotate slower than the gas above this height. Because \textsc{phantom} is a 3D code, our calculations systematically underestimate the 2D analytic solution, which assumes all of the dust is rotating in the mid-plane of the disc. We can reduce this offset by only considering particles near the mid-plane, but having fewer particles to average can make the data more noisy. In making \cref{fig:radial_drift}, we used $10^7$ particles in the disc, and discard all particles with $z > H_\text{g}$ ($\sim1/3$ of the particles). Even with so many particles remaining, the inner $\sim20$ bins are very noisy (note the first 14 are not shown), with values ranging between $-1.5$ and $0.3 \, \eta v_\text{K}$. Also note that the standard deviation for most bins is larger than the size of the plotting window, with typical magnitudes ranging from tens to hundreds $[\eta \vK]$. Thus we should not take the discrepancy between the numerical and analytic solutions too seriously.

\begin{table}
	\centering
	\caption{The initial grain sizes ($s$), dust fractions ($\epsilon$), and dust-to-gas ratios ($\mathcal{E}$) used in the outward migration test in \cref{fig:radial_drift}. Variables with/without the subscript $j$ indicate \textsc{multigrain}/single-phase values, respectively. The final row gives the sum of the individual dust fractions and dust-to-gas ratios, highlighting the inherent discrepancies between setups of single- and multi-phased simulations.
	}
	\label{tab:dust_vals_migration}
	\sisetup{table-format = 1.2,table-auto-round = true}
	\begin{tabular*}{\columnwidth}
		{@{\extracolsep{\stretch{1}}}
			S[table-format=1.2e-1,scientific-notation=true]
			S[table-format=1.2e-1,scientific-notation=true]
			S[table-format=1.2e-2,scientific-notation=true]
			S[table-format=1.2e-1,scientific-notation=true]
		@{}} \toprule
		{$s$ and $\sj\,$[cm]} 		&	{$\epsilon$ and $\epsj$}	&	{$\mathcal{E}$}		&	{$\mathcal{E}_j$}	\\\midrule
		1.000000000000000E-5	&	1.3421160296E-3	&	1.3439197259E-3	&	2.0131740445E-3	\\
		2.782559402207126E-5	&	2.2390023317E-3	&	2.2440267127E-3	&	3.3585034976E-3	\\
		7.742636826811278E-5	&	3.7352444428E-3	&	3.7492488036E-3	&	5.6028666642E-3	\\
		2.154434690031882E-4	&	6.2313695927E-3	&	6.2704430408E-3	&	9.3470543890E-3	\\
		5.994842503189409E-4	&	1.0395562485E-2	&	1.0504765430E-2	&	1.5593343727E-2	\\
		1.668100537200059E-3	&	1.7342530845E-2	&	1.7648602275E-2	&	2.6013796268E-2	\\
		4.641588833612777E-3	&	2.8931900180E-2	&	2.9793894152E-2	&	4.3397850271E-2	\\
		1.291549665014883E-2	&	4.8266014662E-2	&	5.0713766037E-2	&	7.2399021993E-2	\\
		3.593813663804626E-2	&	8.0520399863E-2	&	8.7571708878E-2	&	1.2078059979E-1	\\
		0.100000000000000E-0	&	1.3432919290E-1	&	1.5517352763E-1	&	2.0149378935E-1	\\\midrule
		{Sum total:}			&	{$1/3$}			&	3.65013902685E-1	&	{$1/2$}			\\\bottomrule
	\end{tabular*}
\end{table}

\section{Discussion and Conclusions}
\label{sec:conclusions}

We have derived and implemented a numerical scheme using SPH that is capable of simulating multiple dust phases composed of small dust grains coupled to the gas in the terminal velocity approximation (i.e. when the stopping time is short compared to the computational timestep). Our method simulates dust using a dimensionless dust fraction, as opposed to traditional methods that employ additional sets of simulation particles. By expanding the scalar dust fraction into an array of $N$ dust fractions that are independently evolved and coupled to the gas, we obtain a method that scales better in terms of computational time and resources as $N$ becomes large. Another benefit of evolving the mixture is that the \textsc{multigrain} method circumvents having to resolve the prohibitive temporal and spatial resolution criteria for small dust grains that usually choke multi-fluid simulations with separate gas and dust particles. 

We have demonstrated that the \textsc{multigrain} continuum and discretised equations correctly reduce to the equations described by \citetalias{Price/Laibe/2015} when $N =1$ and that there is no loss in accuracy when simulating a single phase using our \textsc{multigrain} framework --- even when that dust phase is divided into multiple mass bins. On the other hand, when simulating multiple unique dust phases, the \textsc{multigrain} method is superior to using multiple single-phase simulations, not only in terms of computational speed and efficiency as discussed above, but also in terms of accuracy as a result of capturing the indirect coupling (via the gas) between dust phases. Although the deviations between our \textsc{multigrain} and single-phase simulations were small for the select test cases we performed in \cref{sec:dusty_settle} ($\sim$ few per cent), there are a few additional points to consider for general applications: (i) perturbations from other dust phases accumulate over time, (ii) perturbations from concentrated dust grains (or equivalently, higher dust-to-gas ratios) are larger in magnitude than for dispersed grains, and (iii) perturbations between phases can further be accentuated by motion of the gas (as opposed to the stationary gas phase in our settling tests). In light of these concerns, we caution against using single-phase simulations where possible and encourage the adoption of the more accurate and efficient \textsc{multigrain} method we present here.

Finally, the present \textsc{multigrain} algorithm can only be used for small dust grains within the terminal velocity approximation, which is accurate only when the stopping time is shorter than the dynamical timescale \citepalias{Laibe/Price/2014a}. To extend to larger grains, we would need to either implement the full multiphase one-fluid equations with implicit timestepping from \citetalias{Laibe/Price/2014c} or develop a hybrid between the one- and multi-fluid methods. Both have advantages and disadvantages, but are beyond the scope of this paper. Presently, we do not account for the evolution in grain size through growth and fragmentation. However, incorporating grain size evolution into the \textsc{multigrain} framework would be straightforward because the mass and number of the simulation particles does not have to change with time.

\section*{Acknowledgments}
We thank the anonymous referee whose careful review helped improve this paper significantly. We would also like to acknowledge Matthew Bate, Giovanni Dipierro, and Christophe Pinte for useful discussions. DJP is grateful for funding via an Australian Research Council (ARC) Future Fellowship, FT130100034. MAH and DJP were funded by ARC Discovery Project grant DP130102078. GL acknowledges financial support from PNP, PNPS, PCMI of CNRS/INSU, CEA and CNES, France. This work has in part been carried out within the framework of the National Centre for Competence in Research PlanetS, supported by the Swiss National Science Foundation. Finally, computations have been done on both the SwinSTAR supercomputer hosted at Swinburne University of Technology and the Piz Daint supercomputer hosted at the Swiss National Computational Centre.

\bibliography{$HOME/Dropbox/Bibtex_library/library}


\appendix

\section{Solutions to the settling test}
\label{sec:settling_test_derivation}

In the limit of very small dust-to-gas ratios, we can neglect the backreaction of the dust on the gas. The dust can then be treated as $N$ independent phases, moving inside a static gas background, and governed by the following one-dimensional equations
\begin{align}
	\frac{\partial \rhod}{\partial t} + \vdone \frac{\partial \rhod}{\partial z} & =  - \rhod \frac{\partial \vdone}{\partial z},
	\label{eq:rhod_eom}
\\
	\frac{\partial \vdone}{\partial t} + \vdone \frac{\partial \vdone}{\partial z} & =  - \frac{\vdone}{\tsone} - \frac{\mathcal{G} M z}{\left( r^2 + z^2 \right)^{3/2}},
	\label{eq:vd_eom}
\end{align}
where we have dropped the subscript $j$ to emphasise that the phases are no longer coupled. To aid our analysis, we define the dimensionless variables $\bar{\rho} \equiv \rho/\rhogO$, $\bar{v} \equiv v/\vK$, $\bar{z} \equiv z/r$, and $\bar{t} \equiv t \, \OmegaK$, where $\vK = \sqrt{\mathcal{G} M/r}$ and $\OmegaK = \vK/r$ are the Keplerian velocity and frequency, respectively. Substituting these quantities into \cref{eq:rhod_eom,eq:vd_eom}, we obtain the corresponding non-dimensionalised equations in the form
\begin{align}
	\frac{\partial \barrhod}{\partial \bart} + \barvd \frac{\partial \barrhod}{\partial \barz} & =  - \barrhod \frac{\partial \barvd}{\partial \barz},
	\label{eq:rhod_nodim}
\\
	\frac{\partial \barvd}{\partial \bart} + \barvd \frac{\partial \barvd}{\partial \barz} & =  - \frac{\barvd}{\St} - \frac{\barz}{\left(1 + \barz^2 \right)^{3/2}},
	\label{eq:vd_nodim}
\end{align}
where $\St = \tsone \OmegaK$ is the Stokes number.

\subsection{Analytic solution to settling problem}
\label{sec:settling_analytic}

Importantly, \cref{eq:vd_nodim} is independent of $\barrhod$. As a first order partial differential equation, a solution for $\barvd$ could potentially be obtained via the method of characteristics. We took a simpler approach by solving the Lagrangian form of the equation with a convective derivative. In this form, the equation is a first order ordinary differential equation that can be solved using an integrating factor. Assuming the dust starts from rest, the solution for the dust velocity is
\begin{equation}
	\barvd = \frac{\St \, \barz}{\left(1+\barz^2\right)^{3/2}} \left( \mathrm{e}^{-\bart/\St} - 1\right).
	\label{eq:vd_solution}
\end{equation}

The same procedure can be used to obtain an equation for the dust density. However, the resulting solution does not conserve mass since it assumes that an infinitely extended dust distribution continuously rains down onto the disc. The density solution nevertheless correctly predicts the location of the incoming dust front and also the interior density profile so long as $t$ is less than the settling timescale. This is not a problem in the velocity solution because all of our dust grains settle at their terminal velocity within the disc, effectively erasing any built up momentum gained at higher altitudes.

\subsection{1D numerical solution to the settling problem}
\label{sec:settling_numeric}

To remove the assumptions imposed by the analytic solution, we also compared our {\sc multigrain} results with a numerical solution to \cref{eq:rhod_nodim,eq:vd_nodim}. We solved the equations on a one dimensional grid using an implicit Crank-Nicolson algorithm, using forward differences in time and centred differences in space. Because the temporal derivative is centred half a timestep in the future, we replace all of the other terms with time averages centred about the same time. Grouping terms based on their location in time, we can then write each equation as a linear system in the form
\begin{equation}
	\mathbf{A} \mathbf{x}^{n+1} = \mathbf{B} \mathbf{x}^n + \mathbf{C},
\end{equation}
where superscripts designate the time level, $\mathbf{A}$ and $\mathbf{B}$ are square sparse matrices, $\mathbf{x}$ is the fluid variable for which we are solving, and the column vector $\mathbf{C}$ is a placeholder for all terms independent of $\mathbf{x}$ ($\mathbf{C}=0$ when $\mathbf{x}$ represents density). Once the boundary conditions have been accounted for in $\mathbf{A}$ and $\mathbf{B}$, the solution at time $n+1$ can be obtained symbolically via
\begin{equation}
	\mathbf{x}^{n+1} = \mathbf{A}^{-1} \left( \mathbf{B} \mathbf{x}^n + \mathbf{C} \right).
\end{equation}
The nonlinearity in the advection term in \cref{eq:vd_nodim} keeps us from obtaining the solution using the exact method as outlined above because it would irreversibly mix terms from different timesteps. To overcome this problem, we assume the leading $\barvd$ in the advection term is known and designate it as $\tildevd$ to keep it separate from the other velocity terms. We account for $\tildevd$ and the fact that the fluid equations are coupled by using a predictor-corrector scheme to advance the system forward in time. Designating predicted quantities with asterisks, we advance the system in four steps: 
\begin{enumerate}[leftmargin=1\parindent]
	\item $\barrhod^*$ is predicted assuming $\barvd$ is constant (i.e. $\barvd^{n+1} = \barvd^n$),
	\item $\barvd^*$ is predicted assuming that $\tildevd$ is constant and equal to $\barvd^n$,
	\item $\barrhod^{n+1}$ is corrected assuming $\barvd^{n+1} = \barvd^*$,
	\item $\barvd^{n+1}$ is corrected assuming $\tildevd^{n+1} = \barvd^*$ and $\tildevd^n = \barvd^n$.
\end{enumerate}
 
Using the same physical parameters as in \cref{sec:dusty_settle}, we discretise the region $z \in [-3H,3H]$ with 1002 cell-centred grid points, including ghost points. The boundary condition for the velocity is $\vdone(\pm3H,t)=0$, which consequently enforces the following boundary condition for the density:
\begin{equation}
	\frac{\partial \barrhod}{\partial \bart} + \barrhod \frac{\partial \barvd}{\partial \barz} = 0,
\end{equation}
at the same locations. The initial conditions are $\vdone(z,0) = 0$ and $\rhod(z,0) = \epsj \rhog$. We found that a dimensionless timestep of 1 was sufficient to keep the algorithm stable. As the dust settles, we do get some low-density numerical noise in the wings of the disc, but this noise is always separated from the settling dust layer by a region of zero density. We have verified that our results do not change when we force the density to zero beyond the first encountered zero-density grid point on either side of the mid-plane.

The close match between our \textsc{multigrain} results and the analytic and numerical solutions demonstrates that our method works.

\label{lastpage}
\end{document}